\documentclass[natbib]{svjour3}                     
\smartqed  
\usepackage{graphicx}
 \usepackage{aps-bibstyle}  
\newcommand{\ion}[2]{#1\,{\sc{#2}}}
\newcommand {\bc}{\begin {center}}
\newcommand {\ec}{\end {center}}
\newcommand {\be}{\begin {equation}}
\newcommand {\ee}{\end {equation}}
\newcommand{\xmm}{{\em XMM-Newton}}
\newcommand{\arcmin}{$^{\prime}$}

\begin{document}
\title{Resonant scattering of X-ray emission lines in the hot
  intergalactic medium 
\thanks{This work was supported by the DFG grant CH389/3-2, grant
  NSh-5069.2010.2, grant RFFI 09-02-00867 and by programs P-04 and
  OFN-16 of the Russian Academy of Sciences. IZ thanks the International
 Max Planck Research School (IMPRS) in Garching. SS acknowledges
 the support of the Dynasty Foundation. } 
} 



\author{Eugene~Churazov \and Irina~Zhuravleva \and Sergey~Sazonov \and
  Rashid~Sunyaev.} 


\institute{E. Churazov \at
Max Planck Institute for Astrophysics, Karl-Schwarzschild str. 1, Garching, 85741, Germany\\
and Space Research Institute, Profsoyuznaya str. 84/32, Moscow,
  117997, Russia
\and 
I. Zhuravleva \at
Max Planck Institute for Astrophysics, Karl-Schwarzschild str. 1, Garching, 85741, Germany
\and
Sergey Sazonov \at
   Space Research Institute, Profsoyuznaya str. 84/32, Moscow,
  117997, Russia  
\and 
Rashid Sunyaev \at
Max Planck Institute for Astrophysics, Karl-Schwarzschild str. 1, Garching, 85741, Germany\\
and Space Research Institute, Profsoyuznaya str. 84/32, Moscow,
  117997, Russia 
}

\date{Received: date / Accepted: date}

\maketitle

\begin{abstract}
  While very often a hot intergalactic medium (IGM) is optically
  thin to continuum radiation, the optical depth in resonant lines can be of
  order unity or larger. Resonant scattering in the brightest X-ray
  emission lines can cause distortions in the surface brightness
  distribution, spurious variations in the abundance of heavy
  elements, changes in line spectral shapes and even polarization
  of line emission. The magnitude of these effects not
  only depends on the density, temperature and ionization state of the
  gas, but is also sensitive to the characteristics of the gas
  velocity field. This opens a possibility to use resonant scattering as a 
  convenient and powerful tool to study IGM properties. We discuss the
  application of these effects to galaxy clusters.  
\end{abstract}

\section{Introduction}
Hot ($10^6-10^8$K) X-ray emitting astrophysical plasmas are often optically
thin with respect to free-free absorption. One famous example is the
coronae of the Sun and other stars. It was realized early
\citep[e.g.][]{1956ZA.....41...67E} that in the solar corona the
optical depth in certain resonant lines of oxygen, neon and iron can
be larger than unity and a radiative transfer problem has to be 
solved in order to get the correct line ratios
\citep[e.g.][]{1978ApJ...225.1069A,1985ApJ...297..338R}.

Below we concentrate on the resonant scattering in the hot plasma of galaxy
clusters. Hot plasma is the dominant constituent of baryons in
clusters, making 10--15\% of the total cluster mass of
$10^{14}-10^{15}~M_\odot$. Bremsstrahlung, recombination continuum and
emission lines excited by electron collisions are the main emission
components in this plasma. In spite of such a large gas mass, the
optical depth of the hot plasma for free-free absorption is extremely
low in the X-ray regime. This is immediately clear from a comparison
of the observed X-ray luminosity of the brightest clusters, $L_X\sim
10^{45}~{\rm erg~s^{-1}}$, with the black body radiation of a Mpc size
object with a temperature $T\sim 10^7-10^8$K:

\be \tau_{ff}\approx\frac{L_X}{4\pi R^2 \sigma T^4}=1.5\cdot10^{-33} \left
( \frac{L_X}{10^{45}~{\rm erg~s^{-1}}} \right ) \left ( \frac{R}{{\rm
    Mpc}} \right )^{-2} \left ( \frac{T}{10^8~{\rm K}} \right )^{-4},
\ee where $R$ is the characteristic cluster size. The depth for
Thomson scattering is also low: \be \tau_T=\sigma_T n_e R= 2\cdot
10^{-3} \left ( \frac{n_e}{10^{-3} {\rm cm^{-3}}} \right ) \left (
\frac{R}{{\rm Mpc}} \right ), \ee where $n_e$ is the electron
density. This means that effectively clusters are transparent for
X-rays. As in the solar corona, this statement is not valid for the
brightest X-ray resonant lines of the most abundant elements like
iron. \citet{Gil87} have shown that the optical depth in the 6.7 keV
line of He-like iron in the brightest cluster A426 (Perseus) is larger
than unity\footnote{For pure thermal broadening of the line.} and
therefore effects of resonant scattering should be visible in the
surface brightness distribution, element abundance variations or in the
spectral shape of the line. It was immediately clear that all these effects are
sensitive to the turbulent broadening of the line, making resonant
scattering a unique tool to probe the characteristic amplitude of the
gas velocities using telescopes with limited energy resolution. The
impact of resonant scattering on surface brightness profiles
has been studied via radiative transfer calculations 
\citep[e.g.][]{Gil87,1998ApJ...497..587S,2004MNRAS.347...29C,zhu10b}
and searched for in observations of clusters, groups of galaxies and
individual massive elliptical galaxies.   

It has been realized that the scattering should also lead to a polarization of
the same lines \citep{Saz02,Zhu10a}, with the polarization degree
varying from $\sim 0$ at the center of a cluster to tens of per cent
in the cluster outskirts (although the unpolarized continuum and nearby
unpolarized lines will decrease this figure for an instrument with
modest spectral resolution). 

There have been numerous experimental studies aimed at detecting effects of
resonant scattering in rich clusters and elliptical galaxies. For the
Perseus cluster, an initial analysis suggested that resonant scattering
effects were present \citep{1998ApJ...499..608M,1999AN....320..283A},
although subsequent 
studies
\citep{2004MNRAS.347...29C,2004ApJ...600..670G}
did not support these results \citep[see
  also][]{2001PASJ...53..595E,2001ApJ...548..141D,2004MNRAS.349..952S}. 
Several other clusters, including M87, Centaurus and A2199, were searched 
for resonant scattering effects 
\citep[e.g][]{1999NuPhS..69..567K,1999AN....320..283A,2000AdSpR..25..603A,2006MNRAS.370...63S,2001ApJ...550L..31M,2002A&A...391..903S}, with mixed results. For
M87 and Perseus, resonant scattering was mentioned as one of the possible
explanations of the central abundance dip
\citep{2001A&A...365L.181B,2004MNRAS.349..952S}.  

In one object -- the elliptical galaxy NGC4636 -- most studies
\citep{2002ApJ...579..600X,2003ASPC..301...23K,2009MNRAS.398...23W,2009PASJ...61.1185H} 
agree that resonant scattering modifies the flux ratio of the Fe XVII
lines at 15.01 \AA ~and 17.05 \AA. The same study also found evidence,
albeit at low statistical confidence, for resonant
scattering in several other systems: NGC1404, NGC5813 and NGC4472.

In addition to the above mentioned studies several other ideas related
to resonant scattering in hot cluster plasma have been proposed. 
\citet{1988ApJ...335L..39K} \citep[see
  also][]{1989ApJ...345...12S,2006ApJ...643L..73M} suggested to use
measurements of the optical depth of the cluster plasma in lines
together with the X-ray surface brightness to determine both the gas
density and the linear size of the cluster. The known linear size can
then be used for a cosmological distance--angular diameter test.
\cite{2002A&A...393..793S} pointed out that scattering of 
continuum emission in a resonant line can be used to constrain the
past X-ray luminosity of AGN in clusters of galaxies. 
Effects of resonant scattering of the cosmic X-ray background (CXB)
were suggested to play a role in the line emissivity of the WHIM
-- warm-hot intergalactic gas \citep{2001MNRAS.323...93C}, provided that one
 can resolve a substantial fraction of point sources
contributing to the CXB.

The structure of this paper is as follows:
in section \ref{sec:tau} we discuss the most promising resonant lines and
calculate the optical depths for several objects; the impact of 
resonant scattering on line surface brightness profiles and on 
abundance measurements is discussed in section \ref{sec:sb}; in
section \ref{sec:vel} we analyze the sensitivity of the resonant
scattering effects to the gas velocity field; in section \ref{sec:pol}
the polarization of X-ray lines is discussed; in section
\ref{sec:line} we illustrate the effects of resonant scattering on the
line shape; in section \ref{sec:ext} we briefly discuss the possible uses of
resonant scattering for cosmological tests (\S\ref{sec:h0}), for
searches of powerful AGN outbursts in the past (\S\ref{sec:agn}) and
for studying the WHIM (\S\ref{sec:whim}).  

\section{Basics of resonant scattering}
\subsection{Optical depth}
\label{sec:tau}
The cross section for scattering at the center of a resonant line can
be written as
\be
\sigma_0=\frac{\sqrt{\pi}hr_e cf}{\Delta E_D},
\label{eq:sig0}
\ee where $r_e$ is the classical electron radius and $f$ is the
absorption oscillator strength of a given atomic transition and
$\Delta E_D$ is the Doppler width of the line. In a plasma with a
temperature typical of galaxy clusters, the line width is
determined by the velocities of thermal and turbulent motions, rather
than by the radiative width. For example, for the 6.7
keV line of He-like iron the radiative width is $\sim$ 0.3 eV, while
thermal broadening is $\sim$ 3 eV for a 5 keV gas. The Doppler width
of the line is defined as  
\be
\Delta E_D=E_0\left[\frac{2kT_e}{Am_p c^2}+\frac{V_{turb}^2}{c^2}\right]^{1/2},
\label{eq:den}
\ee where $A$ is the atomic
mass of the corresponding element, $m_p$ is the proton mass and
$V_{turb}$ is the characteristic turbulent velocity. $V_{turb}$ is often
parametrized as $V_{turb}=c_{s} M$, where $M$ is the Mach
number\footnote{This form of parametrization is rather arbitrary. For
  isotropic turbulence in a gas with $\gamma=5/3$, the value of $M=1$
  corresponds to the energy in turbulent motions being equal to 0.83 times
  the thermal energy density $\displaystyle \frac{1}{\gamma-1}
  \frac{\rho}{\mu m_p}kT$.} and the
sound speed in the plasma is $c_{s}=\sqrt{\gamma k T/\mu m_p}$, where
$\gamma=5/3$ is the adiabatic index for an ideal mono-atomic gas and
$\mu=0.61$ is the particle mean atomic weight. We can rewrite
the previous expression as 
\be \Delta
E_D=E_0\left[\frac{2kT_e}{Am_p c^2}(1+1.4AM^2)\right]^{1/2}.  
\ee 
The optical depth of the cluster at the center of the line is then 
$\displaystyle \tau=\int {\sigma_0 n_i dl}$,  
where $l$ is the distance along the photon propagation direction
and $n_i$ is the number density of ions in the ground state of a given
transition. For cluster conditions all ions are in the ground state
(the frequency of collisions or any other excitation process are negligible
compared to the radiative decay rate of the excited state) and $n_i=n_p
Z \delta_i$, where $n_p$ is the density of protons, $Z$ the
abundance of the element relative to hydrogen and $\delta_i$ is the
fraction of the element in the appropriate ionization state. That is 
 \be
\tau=\int {\frac{\sqrt{\pi}hr_e cf}{E_0\left[\frac{2kT_e}{Am_p c^2}(1+1.4AM^2)\right]^{1/2}} n_p Z \delta_i dl}.
\label{eq:tau}
\ee 
Obviously, to have a large optical depth in the line one needs an
astrophysically abundant element, a large oscillator strength of the
transition, an appreciable ionization fraction of a given ion and a small line
width. If the line width is dominated by thermal broadening then
the lines of the heaviest elements have an advantage over lighter
elements, since the thermal broadening scales as $\displaystyle 
1/\sqrt{A}$. 

\citet{Gil87} give a convenient
expression for the optical depth (from the cluster center to the
observer) at the midpoint of a Doppler-broadened resonance line when
the isothermal gas density distribution can be approximated by a
$\beta$-model $\displaystyle n=N_0\left [1+\left (\frac{r}{r_c}\right
  )^2 \right ] ^{-3/2\beta}$:
\begin{eqnarray}
\tau_0=\frac{\sqrt{\pi}}{2}
\frac{\Gamma(3\beta/2-1/2)}{\Gamma(3\beta/2)}N_{z,0} r_c\sigma_0
\approx 2.7\frac{\Gamma(3\beta/2-1/2)}{\Gamma(3\beta/2)}
\frac{N_0}{10^{-3}\,{\rm cm}^{-3}}\frac{Z}{Z_\odot}
\,\delta_i(T)
\nonumber \\
\times\frac{r_c}{250\,{\rm kpc}}\frac{\sigma_0(10^7\,{\rm K}, {\mathcal M}=0)}
{10^{-16}\,{\rm cm}^{2}}\left[\frac{T}{10^7\,{\rm K}}
(1+1.4 A{\mathcal M}^2)\right]^{-1/2}
,
\label{eq:tau_0}
\end{eqnarray}
where $\Gamma$ is the gamma-function; parameters of the 6.7
keV line of FeXXV are used; and $Z/Z_{\odot}$  is the abundance of a given
element relative to the solar abundance of iron.  In Table
\ref{tab:tau} we quote the optical depths of several promising lines
for 3 objects, representing different mass limits -- from an
elliptical galaxy (NGC4636) to a rich cluster (Perseus).  The optical
depths were calculated using the observed temperature and density
radial profiles. Pure thermal broadening of the lines was assumed. 

From Table \ref{tab:tau} it is clear that different lines can be
optically thick in objects having drastically different
masses and temperatures. For objects with $T\ge 3$ keV
(e.g. Perseus cluster) the He-like iron line at 6.7 keV is a likely winner;
while for cool systems (e.g. NGC4636 with $T\sim 0.6$ keV) the Ne-like
iron line at 0.83 keV has the largest optical depth.

\begin{table}
 \centering
  \caption{Oscillator strength, Rayleigh
    scattering weights and optical depth of the most promising X-ray lines for the elliptical
galaxy NGC 4636 and the M87/Virgo and Perseus (A426) clusters. The
optical depths were calculated using the observed temperature and density
profiles and assuming the flat abundance profile $Z=0.5 Z_\odot$
(relative to the Solar abundances of \citet{1989GeCoA..53..197A}), 
collisional ionization equilibrium and pure thermal broadening of the lines.} 
   \begin{tabular}{@{}rcccccc@{}}
   \hline
  Ion & $E,~{\rm keV}$ & $f$ &$w_2$&$\tau$, NGC 4636& $\tau$, Virgo/M87 & $\tau$,
 Perseus\\
 \hline
O VIII & 0.65 & 0.28 &0.5& 1.2 & 0.34 & 0.19\\
Fe XVIII & 0.87 & 0.57 &0.32&1.3& 0.0007 & 1.5$\cdot 10^{-7}$\\
Fe XVII & 0.83 & 2.73 &1&8.8& 0.0005 & 2.8$\cdot 10^{-8}$ \\
Fe XXIII & 1.129 & 0.43&1 & 0.016& 1.03 & 0.16\\
Fe XXIV & 1.168 & 0.245&0.5 & 0.002& 1.12 & 0.73\\
Fe XXV & 6.7 & 0.78 &1& 0.0002& 1.44 & 2.77\\
 \hline
 \label{tab:tau}
 \end{tabular}
 \end{table}

\subsection{Phase function}
\label{sec:phase}
For any transition, the resonant scattering can be represented as a
combination of two processes: isotropic scattering with a weight $w_1$
and Rayleigh scattering with a weight $w_2=1-w_1$ \citep{Ham47,
  Chan50}. The weights $w_1$ and $w_2$ depend on the total angular
momentum $j$ of the ground level and on the difference between the
total angular momenta of the excited and ground levels $\Delta j$
(=$\pm 1$ or $0$). The expressions for the weights are given by
\citet{Ham47}.  For a subset of the most promising lines the $w_2$
values are given in Table \ref{tab:tau}.

If the radiation is initially unpolarized, then the probability
for a photon to scatter into a unit sold angle at an angle
$\theta$ with respect to the incident direction is    
\begin{equation}
P(\mu)=\frac{1}{4\pi}\left[w_1+\frac{3}{4}(1+\mu^2)w_2\right],
\end{equation}
where $\mu=\cos \theta$. 

If polarization has to be accounted for (see \S\ref{sec:pol}), then 
the scattering matrix of resonant scattering can be written as
\begin{eqnarray}
\left ( 
\begin{array}{l}
I'_l \\
I'_r \\
U'
\end{array}
\right )=\frac{1}{4\pi} \left [\frac{1}{2} w_1  \left ( 
\begin{array}{lll}
1 ~& 1 ~& 0  \\
1 ~& 1 ~& 0  \\
0 ~& 0 ~& 0 
\end{array}
\right )+
\frac{3}{2} w_2 \left ( 
\begin{array}{lll}
\mu^2 ~& 0 ~& 0  \\
0 ~& 1 ~& 0  \\
0 ~& 0 ~& \mu 
\end{array}
\right )  
\right ] \left ( 
\begin{array}{l}
I_l \\
I_r \\
U
\end{array}
\right ),
\label{eq:stokes}
\end{eqnarray}
where  $I_l, I_r, U$ are the Stokes parameters as defined by
\citet{Ham47,Chan50}. 

For the changes in the surface brightness (see \S\ref{sec:sb}), the
difference between the isotropic and Rayleigh phase functions is not very
important, while it is crucial if one needs to evaluate the degree of
polarization (see \S\ref{sec:pol}). Isotropic scattering does not
produce polarization. Instead it ``erases'' information on the initial
direction and orientation of the electric vector in the ``memory'' of
the scattered photon. On the contrary, Rayleigh scattering changes the
polarization state of the radiation field.

\section{Impact on surface brightness profile and abundance}
\label{sec:sb}
\begin{figure*}
  \includegraphics[width=0.5\textwidth]{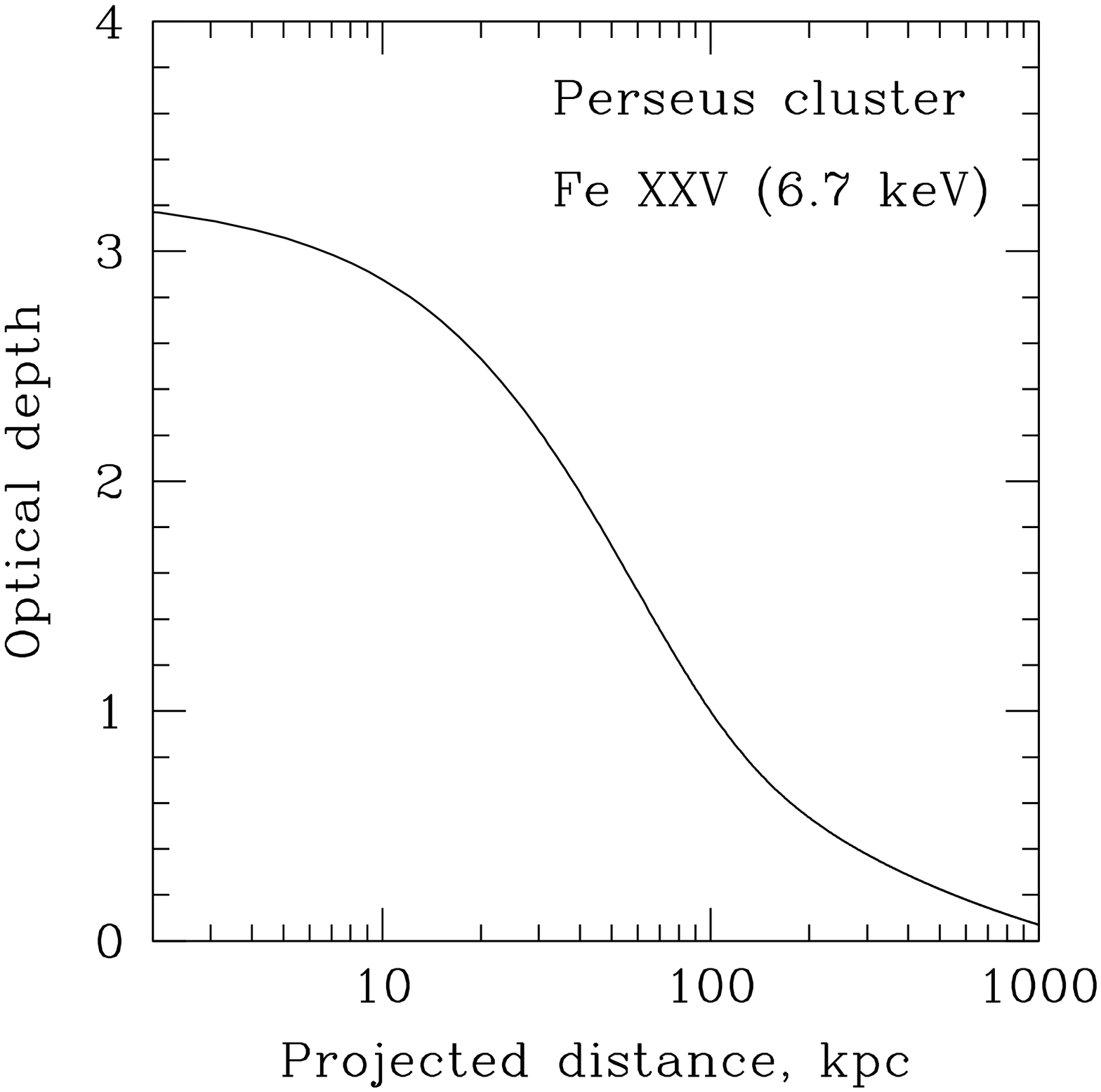}
  \includegraphics[width=0.5\textwidth]{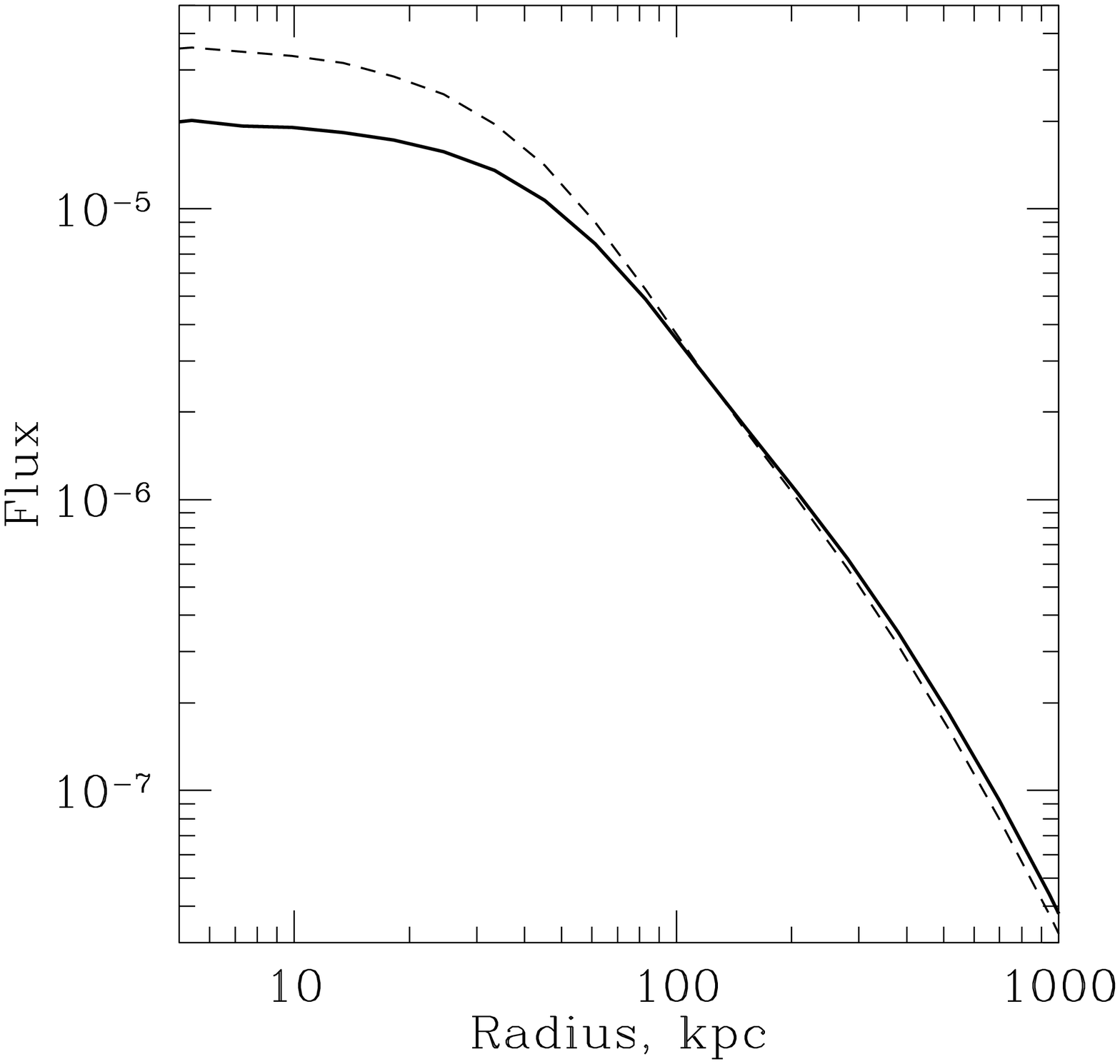}
\caption{{\bf Left:} Optical depth of the Perseus cluster in the 6.7 keV FeXXV
  line over the range of radii from $r$ to $\infty$. For $r_1\approx
  70$ kpc the optical depth is of order 1. To a first
  approximation the surface brightness is expected to be only weakly
  disturbed by resonant scattering for projected radii $R \gg r_1$. For smaller
  projected radii the surface brightness profile is expected to be
  flat.
{\bf Right:} Radial profiles of the He-like iron $K_\alpha$
  line with (thick solid line) and
  without (dashed line) resonant scattering in the
  Perseus cluster. Resonant scattering suppresses the line intensity
  in the core and redistributes line photons to larger radii. Pure
  thermal broadening and a flat radial abundance profile were assumed
  in this calculation. Adapted from \citet{2004MNRAS.347...29C}. 
}
\label{fig:tau}       
\end{figure*}

Since the density of the gas declines faster than $\displaystyle \sim r^{-1}$ in
the outer regions of all clusters, the optical depth of the outer regions in
any line is small (see Fig. \ref{fig:tau}, left). We therefore do not
expect strong modifications of the line surface brightness profile at
large projected distances from the cluster center (see
Fig. \ref{fig:tau}, right). The central part on the contrary is dense
and in the region where the 
optical depth is of order unity one can expect that resonant
scattering will erase all features in the surface brightness, leading
to a flattened (see Fig. \ref{fig:tau}, right) surface brightness
distribution \citep{Gil87}. Due to the conservative nature of resonant
scattering, the photons removed from the line of sight going through
the cluster center are re-distributed to larger projected
distances. This effect causes a slight increase of the surface
brightness  at $R>r_1$ (see Fig. \ref{fig:tau}, right). Since
the surface brightness in the continuum is not modified, the resonant
scattering causes a decrease of the equivalent width of the line in
the cluster center and a slight increase of the equivalent width in
the outer regions. This could cause an apparent decrease in the
abundance of heavy elements in the central part of the cluster if
resonant lines are used to measure it \citep{Gil87}. 

This effect was suggested to play a role in the spurious ``abundance
holes`` in cluster centers
\citep[e.g.][]{2001A&A...365L.181B,2004MNRAS.349..952S}, although
other effects such as the presence of multi-temperature plasma may be
more important in these environments
\citep[e.g.][]{2000ApJ...539..172B}.  

Note that optically thin lines are not modified by resonant
scattering and therefore the observed spectrum might have line ratios
different from the expected values for an optically thin thermal
plasma. This implies that one can use the line ratios (the flux in an
optically thick line divided by the flux in an optically thin line)
rather than the line equivalent width (i.e. the ratio of the
flux in the optically thick line to the spectral intensity of the
continuum). The ideal situation is when two lines from the same ion are
used, since in this case many issues related to the modeling of the
ionization state can be avoided.

A set of simple expressions was derived by \citet{Gil87,Saz02} for
model gas density distributions and in the limit of $\tau \ll 1$ for
the isotropic and Rayleigh phase functions. These expressions can be
used for making order of magnitude estimates, while for real systems
it is necessary to solve a radiative transfer problem to predict the shape 
of distortions, especially when substantial temperature gradients are
present, which affect the ionization state and therefore the radial
distribution of ions. 

Overall the resonant scattering has been searched for in a number of
objects, although clear evidence for the effect was found only in few
systems -- for the Ne-like iron (Fe XVII) line at 15.01 \AA\ in NGC4636 
\citep{2002ApJ...579..600X,2003ASPC..301...23K,2009MNRAS.398...23W,2009PASJ...61.1185H} 
and for NGC1404, NGC5813 and NGC4472
\citep{2009MNRAS.398...23W}. Nevertheless, the magnitude of the effect 
(modification of the surface brightness distribution in the line) is
well within the capabilities of modern X-ray telescopes. In fact,
one can use the absence of resonant scattering effects to infer the 
properties of the gas velocity field, as discussed in the next
section.

\section{Sensitivity to the velocity field}
\label{sec:vel}
\begin{figure*}
  \includegraphics[width=0.5\textwidth]{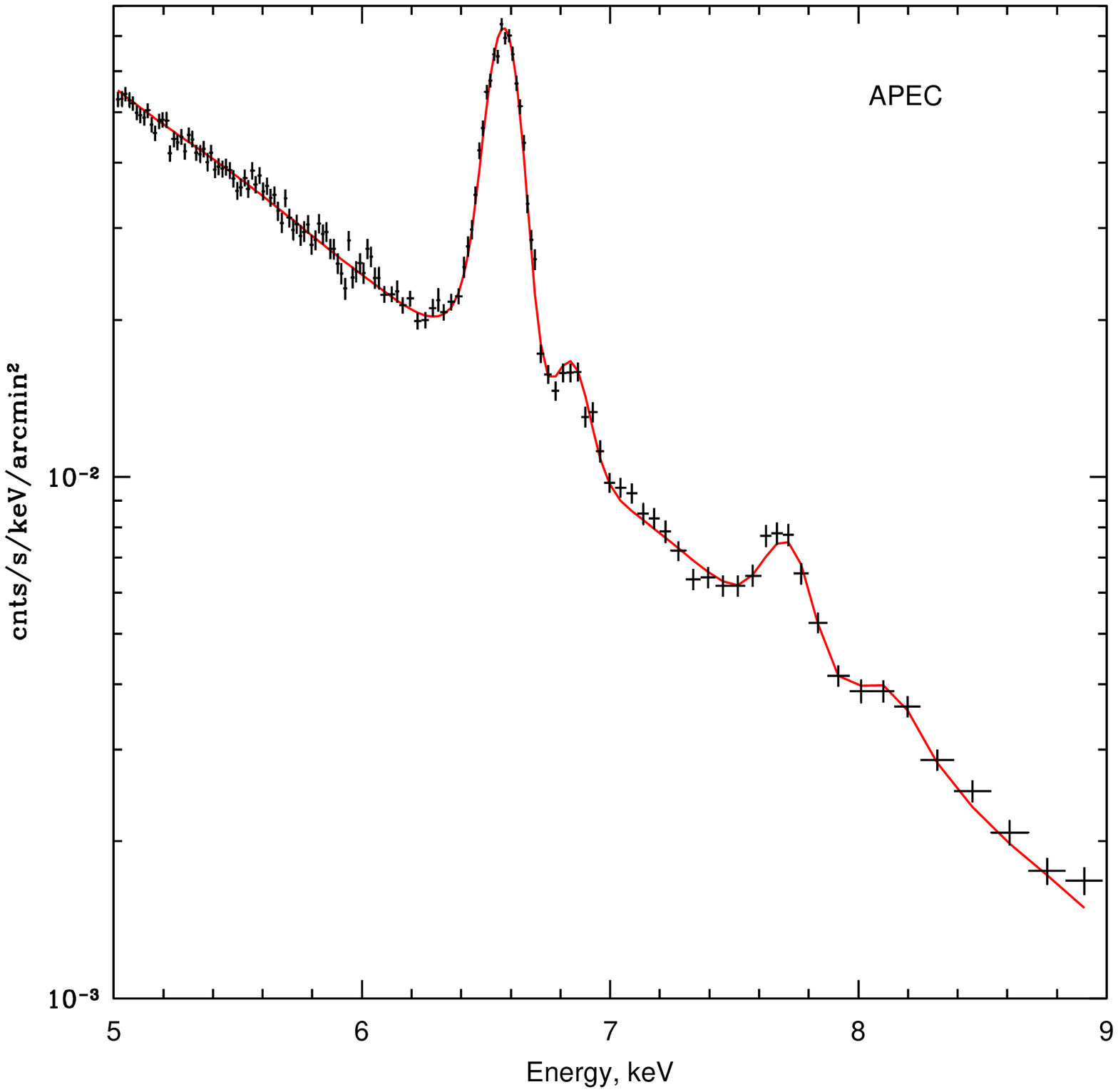}
  \includegraphics[width=0.5\textwidth]{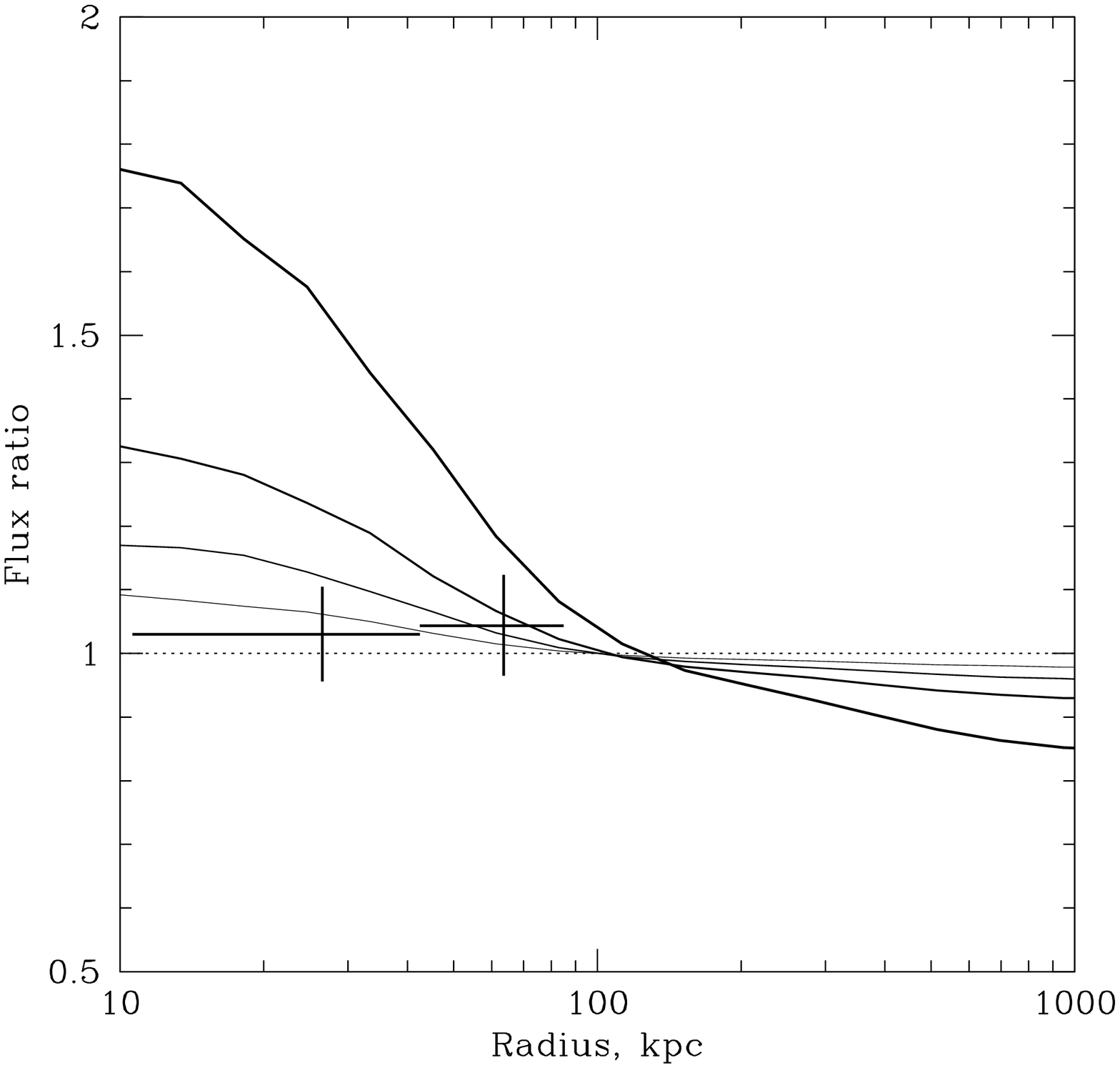}
\caption{{\bf Left:} 5-9 keV spectrum of the $30''$ to
$2'$ annulus centered on NGC~1275 and fitted with APEC \citep{Smi01} models. 
 {\bf Right:} Influence of turbulence on the strength of the resonant
  scattering effect. The curves show the expected ratio of the optically
  thick and optically thin lines calculated 
  for effective Mach numbers of 0, 0.25, 0.5 and 1 (from top to bottom). For
  comparison, crosses show the observed ratio of heavy element abundances
  obtained ignoring parts of the spectrum containing the 6.7 keV and
  7.9 keV complexes, respectively. Given that the resonant $K_\alpha$
  line of He-like iron contributes about 50\% to the 6.6-6.8 keV
  complex of lines, the measured abundance ratio is consistent with
  the curves for Mach number $\ge$0.5.  Adapted from
  \citet{2004MNRAS.347...29C}.} 
\label{fig:chur04}       
\end{figure*}

If one accurately measures the temperature of the intracluster medium
(ICM), then the line ratios of (optically thick) resonant and (optically 
thin) non-resonant lines can be used as a powerful tool to measure the
line widths and therefore to get information on the velocity field, which is
difficult to obtain by other means. Given that the optical depth
sensitively depends on the turbulent broadening of resonant lines,
especially for heavy elements like iron, the comparison of the fluxes
of e.g. iron and nickel lines could be used to place constraints on
the level of turbulence in clusters \citep{2004MNRAS.347...29C}. For
the Perseus cluster, the XMM-Newton spectra of the central region do 
not show a suppression of the resonant 6.7 keV line of iron relative to
a much more optically thin nickel line when the APEC
\citep{Smi01} plasma emission model is used, although one should expect
the 6.7 keV line of He-like iron to be suppressed in the core if the
iron line were only thermally broadened. Indeed, without additional
line broadening the optical depth of the 6.7 keV line is $\sim 3$ (see
Table \ref{tab:tau}). Therefore, the line in the central $\sim 1'$
region must be suppressed by a factor of about 1.7 (see
Fig. \ref{fig:chur04}), in contrast with observations.

\begin{figure*}
\begin{minipage}{0.4\textwidth}
  \includegraphics[width=0.8\textwidth,clip=t,angle=270.]{NGC4636narrow.ps}
\end{minipage}\hspace{7mm}
\begin{minipage}{0.48\textwidth}
  \includegraphics[width=0.95\textwidth]{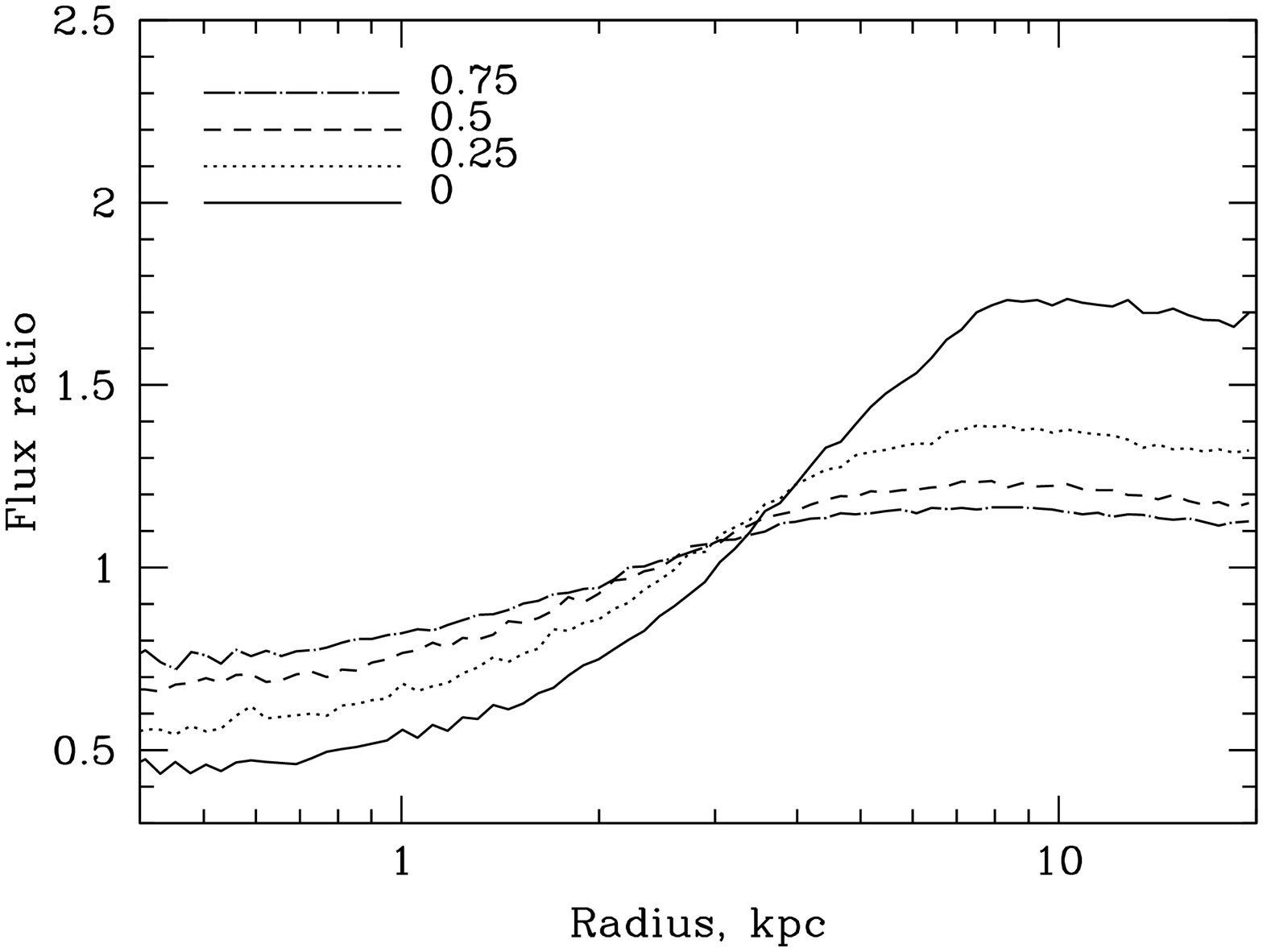} 
\end{minipage}
\caption{{\bf Left:} \xmm\ RGS spectra extracted from a
  0.5\arcmin\ wide region centered on the core of
  NGC~4636. The full line indicates the best optically thin
  single-temperature plasma model fit. The 13.8--15.5~\AA\ part of the
  spectrum, where the strongest \ion{Fe}{xvii} and \ion{Fe}{xviii}
  resonance lines are present, was excluded from these spectral
  fits. {\bf Right:} Simulated radial profiles of the ratio of the
  15.01~\AA\ line intensities calculated with and without the effects
  of resonant scattering, for isotropic turbulent velocities
  corresponding to Mach numbers 0.0, 0.25, 0.5, and 0.75 and a flat abundance
  profile. Adapted from \citet{2009MNRAS.398...23W}.}
\label{fig:wer09}       
\end{figure*}

Even more robust constraints are expected if optically
thick and optically thin lines of the same element or even better of the
same ion are used. \cite{2009MNRAS.398...23W} obtained high
resolution spectra of the giant elliptical galaxy NGC 4636 using the
grating spectrometers on the XMM-Newton satellite. A
detailed study of the spectra proved that the Fe XVII line at 15.01
\AA ~is suppressed 
(Fig. \ref{fig:wer09}) only in the dense core and not in the
surrounding regions, while the line of Fe XVII at 17.05 \AA ~is
optically thin and is not suppressed. \citet{2009MNRAS.398...23W}
modeled the radial intensity profiles of the optically thick line, 
accounting for the effect of resonant scattering for different values of
the characteristic turbulent velocity. Comparing the model to the
data, it was found that the isotropic turbulent velocities on spatial
scales smaller than $\approx$ 1 kpc are less than 100 km/s and the
turbulent pressure support in the galaxy core is smaller than 5 per
cent of the thermal pressure at the 90 per cent confidence level.  

While there is still a bit of a controversy in the results, which
definitely suffer from the limited energy resolution of the
present-day CCDs or from the finite size of the emitting region for
RGS data, it is clear that the magnitude of the effect is within the
reach of the current generation of telescopes and the situation
will improve in future.

\begin{figure*}
  \includegraphics[width=0.5\textwidth]{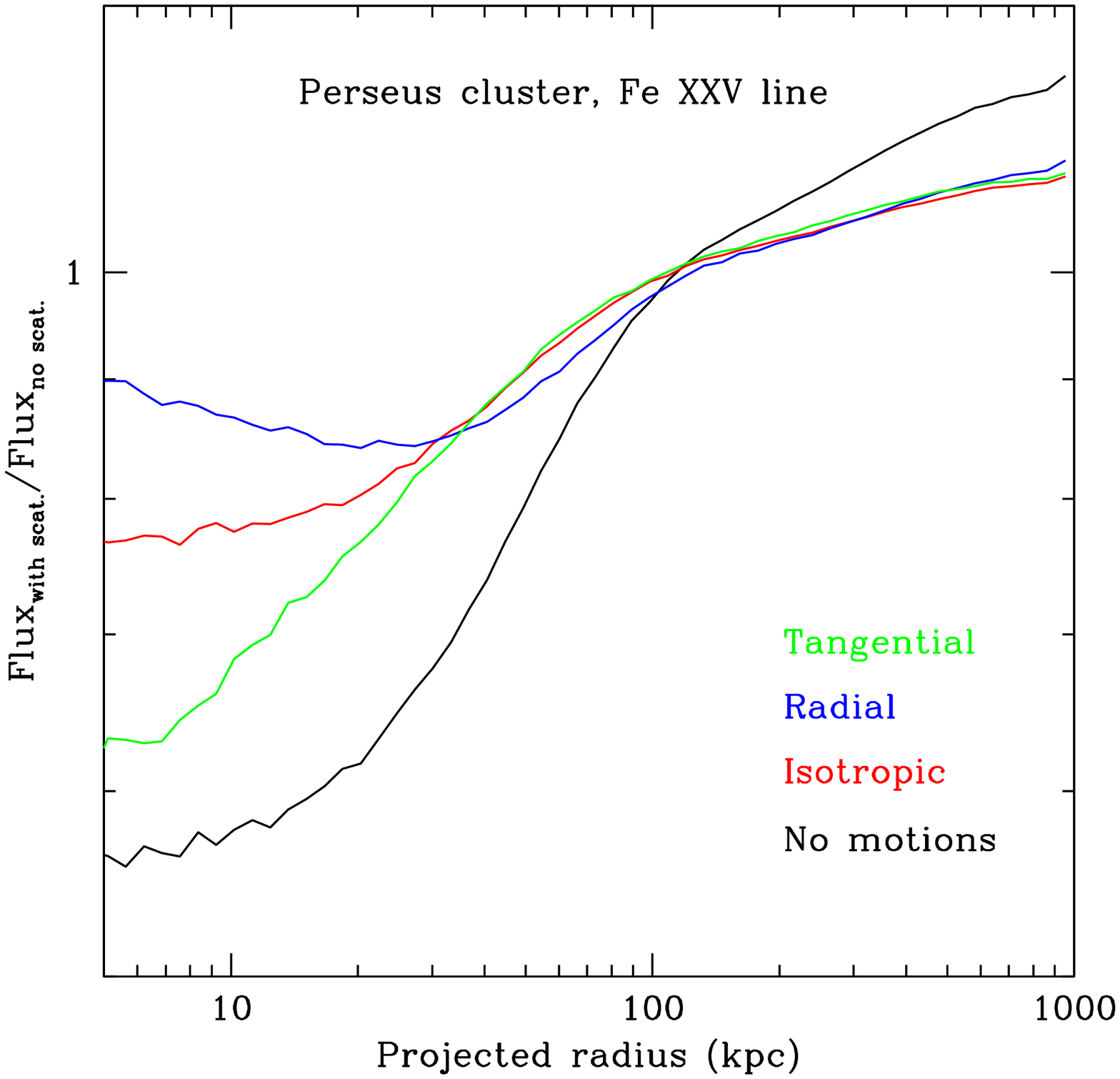}
  \includegraphics[width=0.5\textwidth]{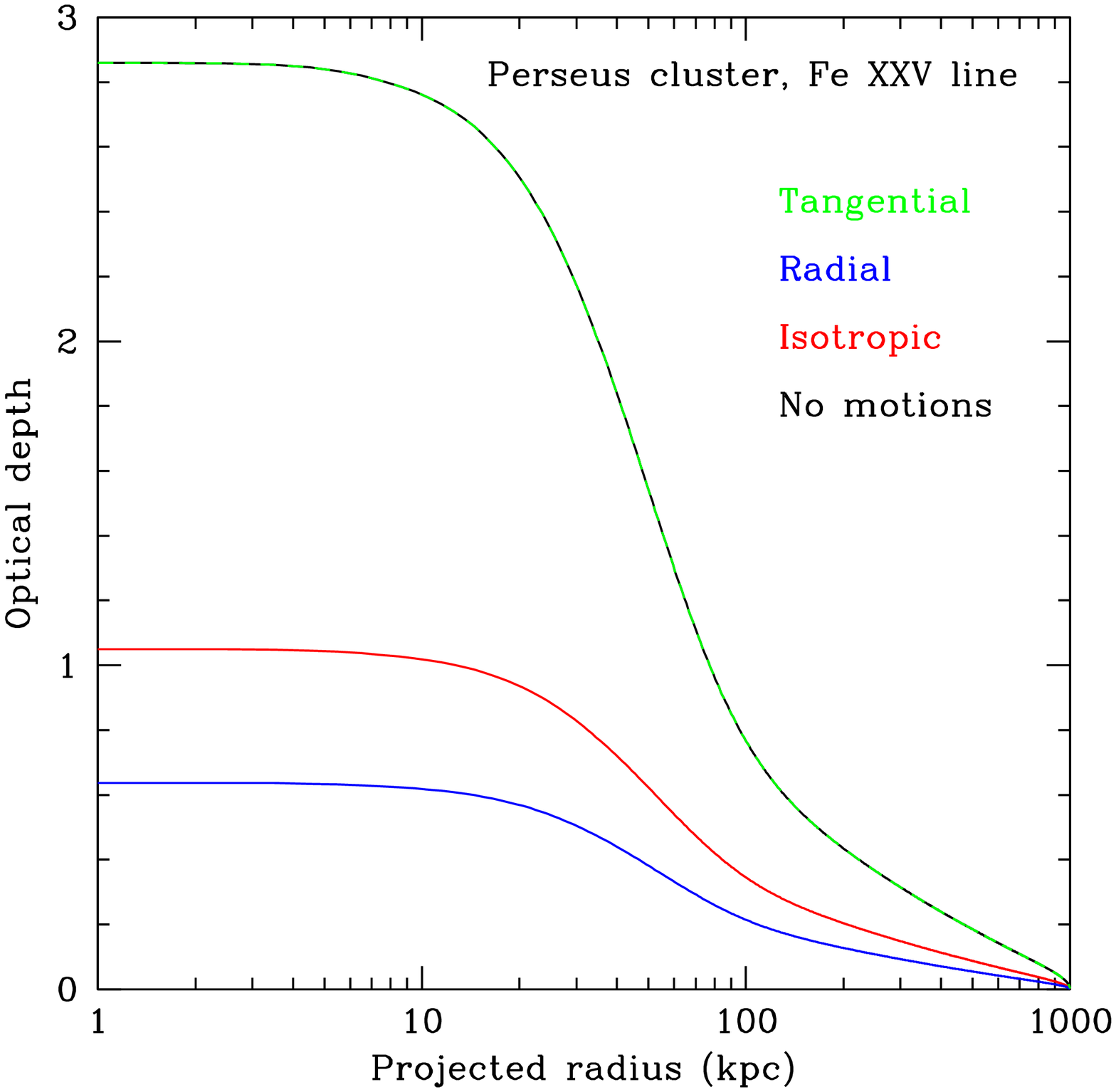}
\caption{{\bf Left:} Impact of the anisotropy of stochastic gas
  motions on the ratio of the fluxes in the Fe XXV line at 6.7 keV in
  the Perseus cluster calculated with and without allowance for scattering. The
  RMS of the 3D velocity field is 500 km/s in all cases (except for
  the case with no gas motions). Adapted from \citet{zhu10b}. 
{\bf Right:} Optical depth of the Perseus cluster in the 6.7 keV FeXXV
line over the range of radii from r to $\infty$, calculated assuming different
directions of stochastic gas motions.
}
\label{fig:ani}       
\end{figure*}

The relation between the resonant scattering effects and the velocity
field can be used to test even more subtle effects, such as
e.g. anisotropy of gas motions. \cite{zhu10b} showed that for pure
radial motions versus pure tangential motions the amplitude of the
effect can change by a factor of 1.5 for realistic values of the
velocity amplitude (see the left panel in Fig. \ref{fig:ani}). Clearly,
tangential gas motions do not change the optical depth in the line (the
right panel, Fig. \ref{fig:ani}) calculated along the radial
direction. Photons coming from the cluster 
central region are expected to make the largest contribution to the scattered
flux. Since they move essentially along the radius, tangential gas
motions do not affect the probablity of these photons to be
scattered. Therefore, to a first approximation the presence of pure
tangential motions does not affect much the magnitude of the resonant
scattering effect. Radial motions, on the contrary, strongly affect
the optical depth for the photons coming from the cluster center (the
right panel, Fig. \ref{fig:ani}) and decrease the efficiency of the
resonant scattering. The net result of the simulations by \cite{zhu10b} is that
the conversion of the observable magnitude of the flux suppression to
the characteristic amplitude of gas motions depends on the anisotropy
of the velocity field. In many situations the assumption of isotropic
stochastic motions is reasonable. But there are cases, e.g. spherical 
shocks/sound waves propagating through the ICM
\citep[e.g.][]{2007ApJ...665.1057F}, when strong anisotropy of the gas 
motions is expected. 

\cite{zhu10b} also tested the impact of the spatial scales of motions on
scattering, showing that suppression of the line flux towards the
center of the cluster is insensitive to the presence of gas motions on
large scales. This is easy to understand since small scale  
motions directly affect the line width and the optical depth of the
line. If gas motions are present on scales larger than the size of
the central region where $\tau$ is of order unity, then these motions
can be interpreted as a motion of the whole central region. The optical
depth and the magnitude of the flux suppression remain unaffected.

\section{Polarization due to resonant scattering}
\label{sec:pol}
\begin{figure*}
  \includegraphics[width=0.5\textwidth]{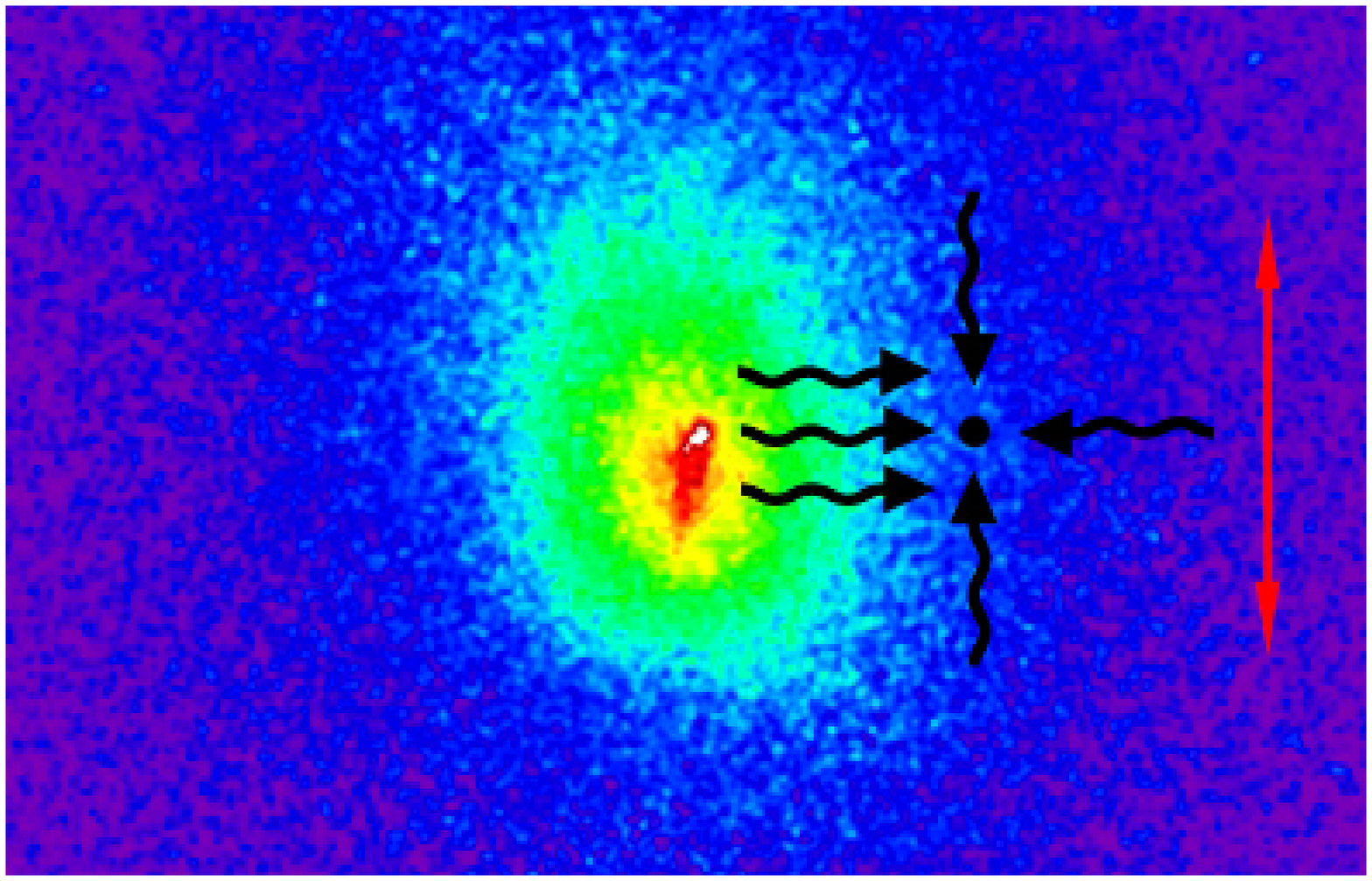}
  \includegraphics[width=0.5\textwidth]{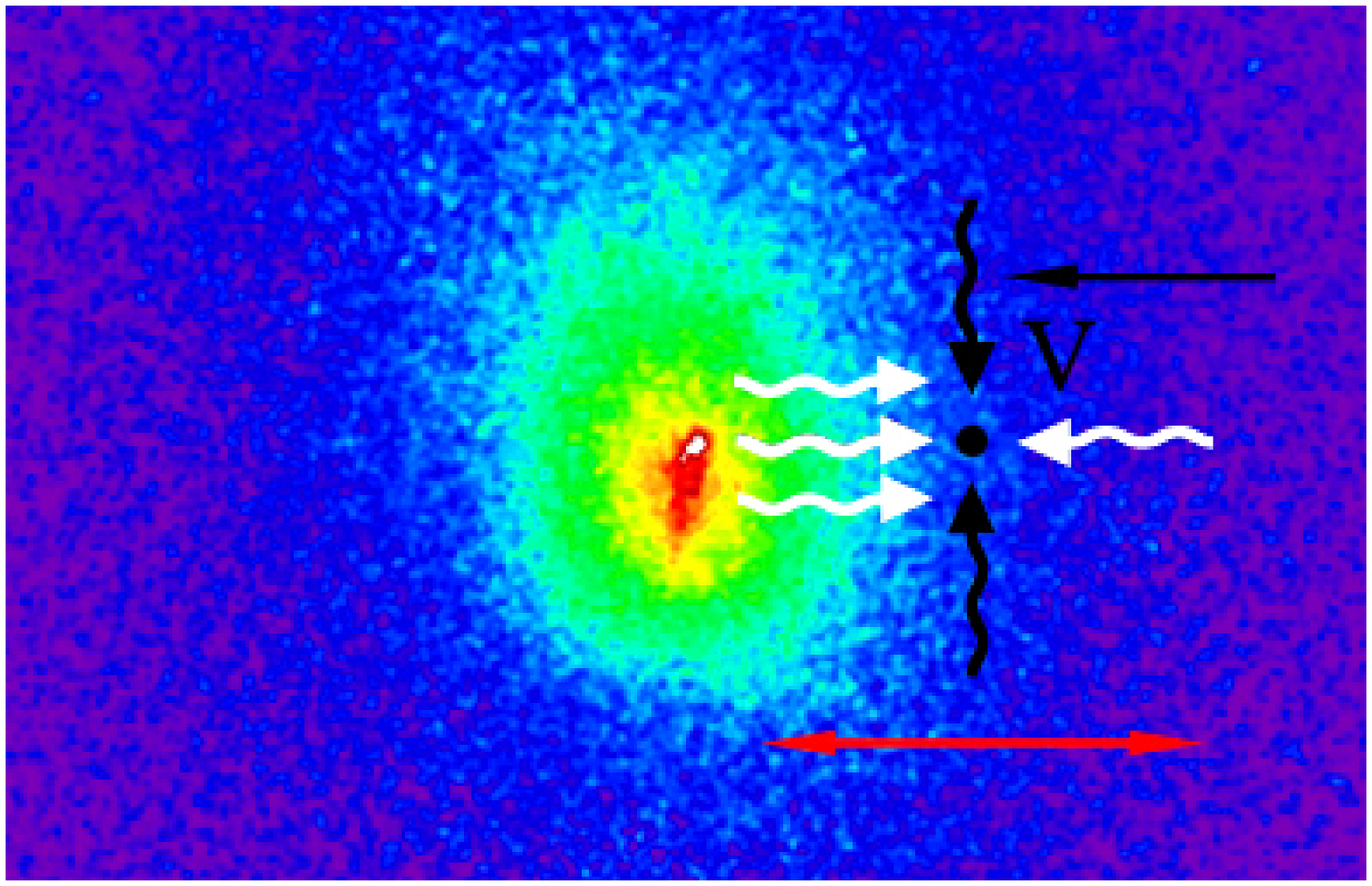}
\caption{Sketch of the problem. Polarization arises due to the
quadrupole component in the unpolarized radiation field produced by a
centrally peaked gas density distribution (the left panel) and/or 
differential gas motions (the right panel). The line of sight is 
perpendicular to the picture plane. {\bf Left panel:} scattering 
occurs at the edge of the cluster (the black dot), the black wavy arrows
show photons coming from different directions and participating in scattering;
the red arrow shows the direction of polarization of the scattered
radiation (the orientation of the electric vector). {\bf Right panel:}
the black dot shows the place where scattering occurs, the black straight
arrow shows the direction of motion of the gas lump with velocity ${\bf V}$
relative to the whole cluster, the black wavy arrows show photons
participating in the scattering while the white wavy arrows show
photons that will not be scattered due to gas motions (the line leaves
the resonance in the direction of motion), the red arrow shows the
direction of polarization of the scattered radiation.} 
\label{fig:geom}       
\end{figure*}

Scattering in strong resonant lines that have a non-zero weight of the
Rayleigh phase function can lead to linear polarization 
of scattered radiation (see section \ref{sec:phase}). To produce net 
polarization during the Rayleigh scattering of initially unpolarized 
light a quadrupole component has to be present in the radiation field. 
It appears quite naturally in galaxy clusters due to (i) the centrally 
peaked gas density distribution and (ii) differential gas motions 
(see Fig. \ref{fig:geom}).

In the cluster center, the radiation field is isotropic (for a spherical
cluster). Therefore, due to the symmetry radiation will be
unpolarized. However, at larger distances from the center a quadrupole
component appears in the radiation field due to the strong dominance of
the flux coming from the cluster center compared to all other
directions. These photons coming from the center give rise to the
polarization of the scattered radiation (see Fig. \ref{fig:geom}, the left
panel). Obviously, the polarization plane is perpendicular to the radius.

\begin{figure*}
  \includegraphics[width=0.5\textwidth]{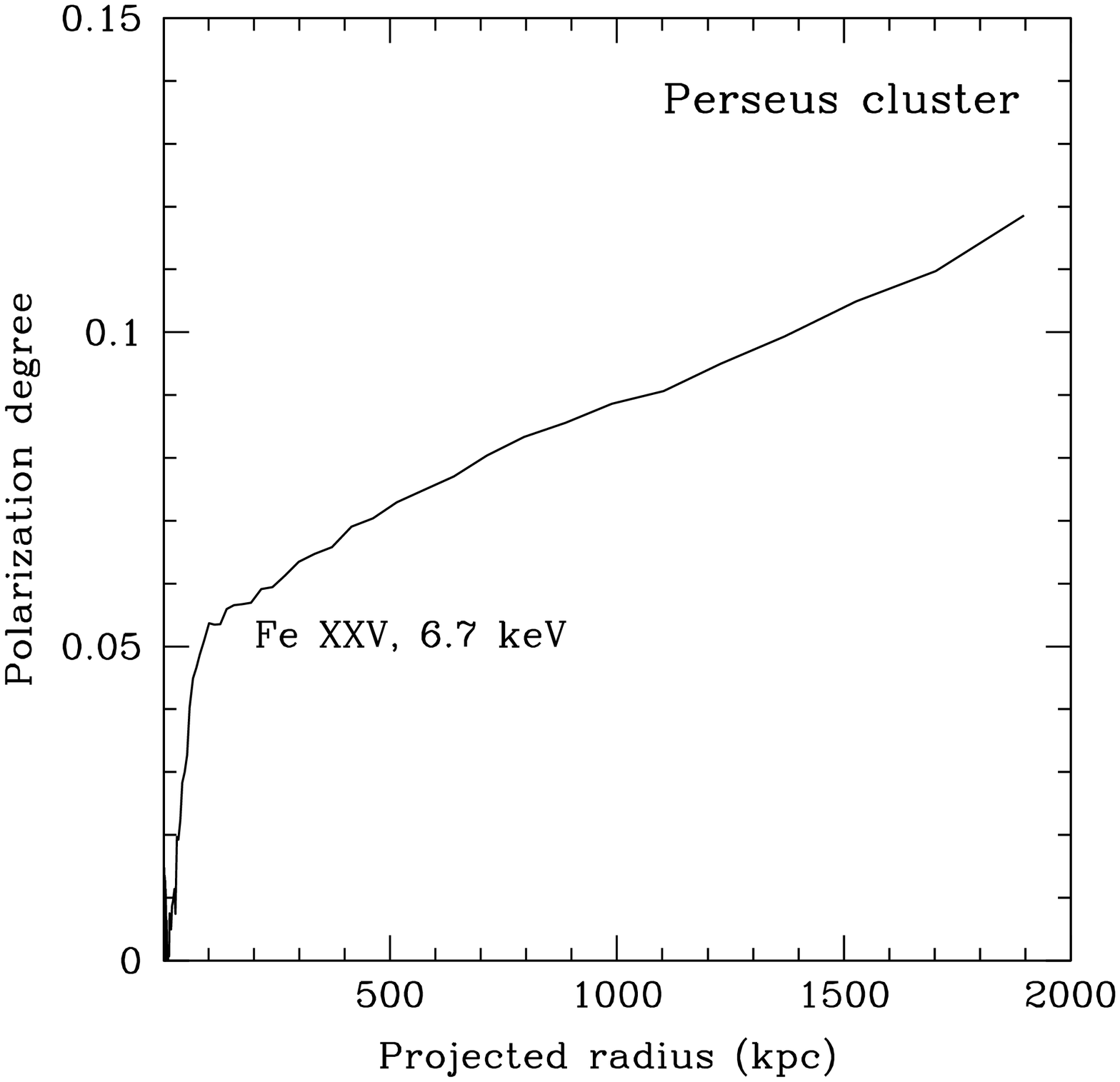}
  \includegraphics[width=0.5\textwidth]{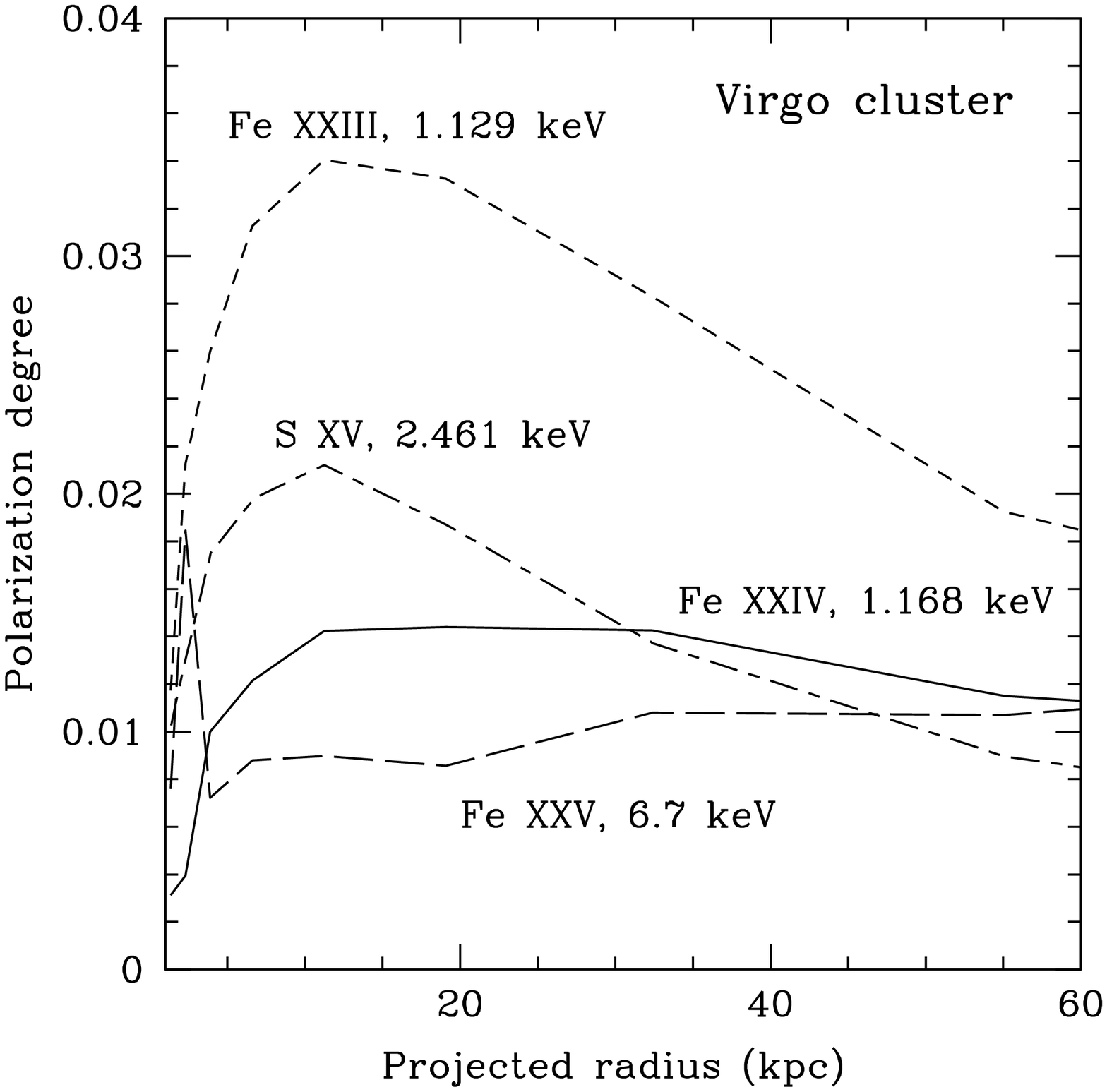}
\caption{{\bf Left panel:} Expected polarization degree as a function
  of projected distance from the center of the Perseus cluster in the
  most prominent resonant line of Fe XXV at 6.7 keV (calculated
  assuming constant iron abundance).  {\bf
    Right panel:} Expected polarization degree as a function of 
projected distance from the center of the Virgo/M87 cluster in the
most prominent resonant lines.}
\label{fig:1Dpol}       
\end{figure*}

The polarization degree was first calculated by \cite{Saz02} 
\citep[see also][]{Zhu10a} for the Perseus and Virgo clusters 
(see Fig. \ref{fig:1Dpol}). For the Perseus cluster, the polarization degree 
was calculated in the He-like iron line at 6.7 keV, which has an optical 
depth $\sim$ 3. The polarization is zero at the center of the cluster 
(as expected from the symmetry of the problem) and increases rapidly 
with distance from the center, reaching $\sim$ 10 per cent at 
$\sim$ 1 Mpc (Fig. \ref{fig:1Dpol}, the left panel). The polarization
degree in the Virgo cluster in the most promising lines is a few per cent 
(Fig. \ref{fig:1Dpol}, the right panel).

The assumption of spherical symmetry is a strong one, since clusters
are rarely spherical.  Depending on the characteristics of the
substructure in the azimuthal density and temperature distributions
the polarization degree can change either way: increase or decrease
even if one neglects non-zero gas velocities.  When working with
simulated clusters we see variations of the polarization degree by as
much as factor of two when 3-dimensional density and temperature
distributions are replaced with the azimuthly averaged values.

Gas motions affect both the amplitude and the direction of
polarization. If a gas lump is moving relative to the whole cluster
with high speed, such that the Doppler shift of the line exceeds the
line width (i.e. the line leaves the resonance), then scattering will
not occur along the direction of motion, while photons coming from the
perpendicular direction will be scattered, producing net polarization.  
The sketch of one of the possibilities is shown of Fig. \ref{fig:geom} 
(the right panel).

\begin{figure*}
  \includegraphics[width=0.5\textwidth]{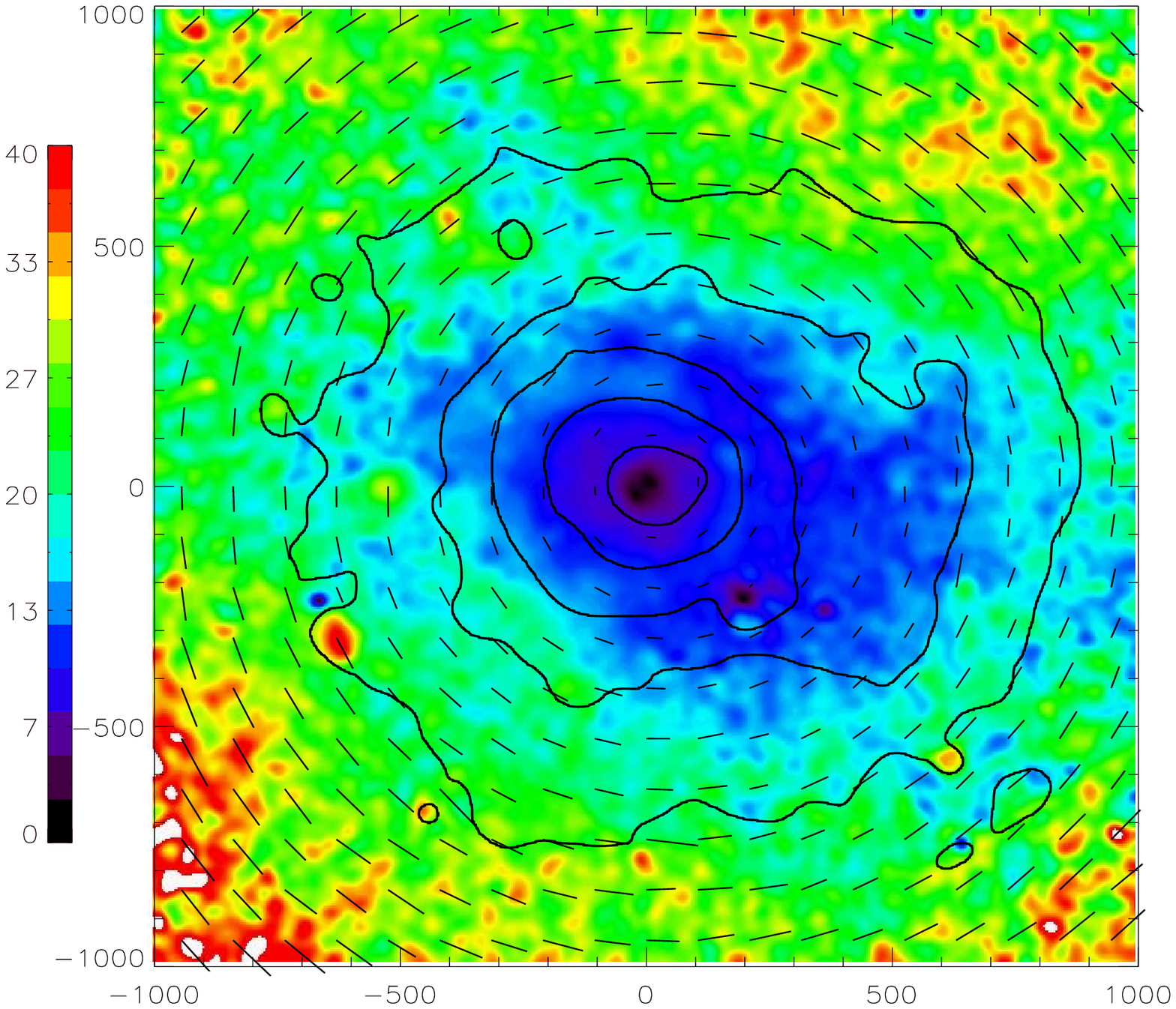}
  \includegraphics[width=0.5\textwidth]{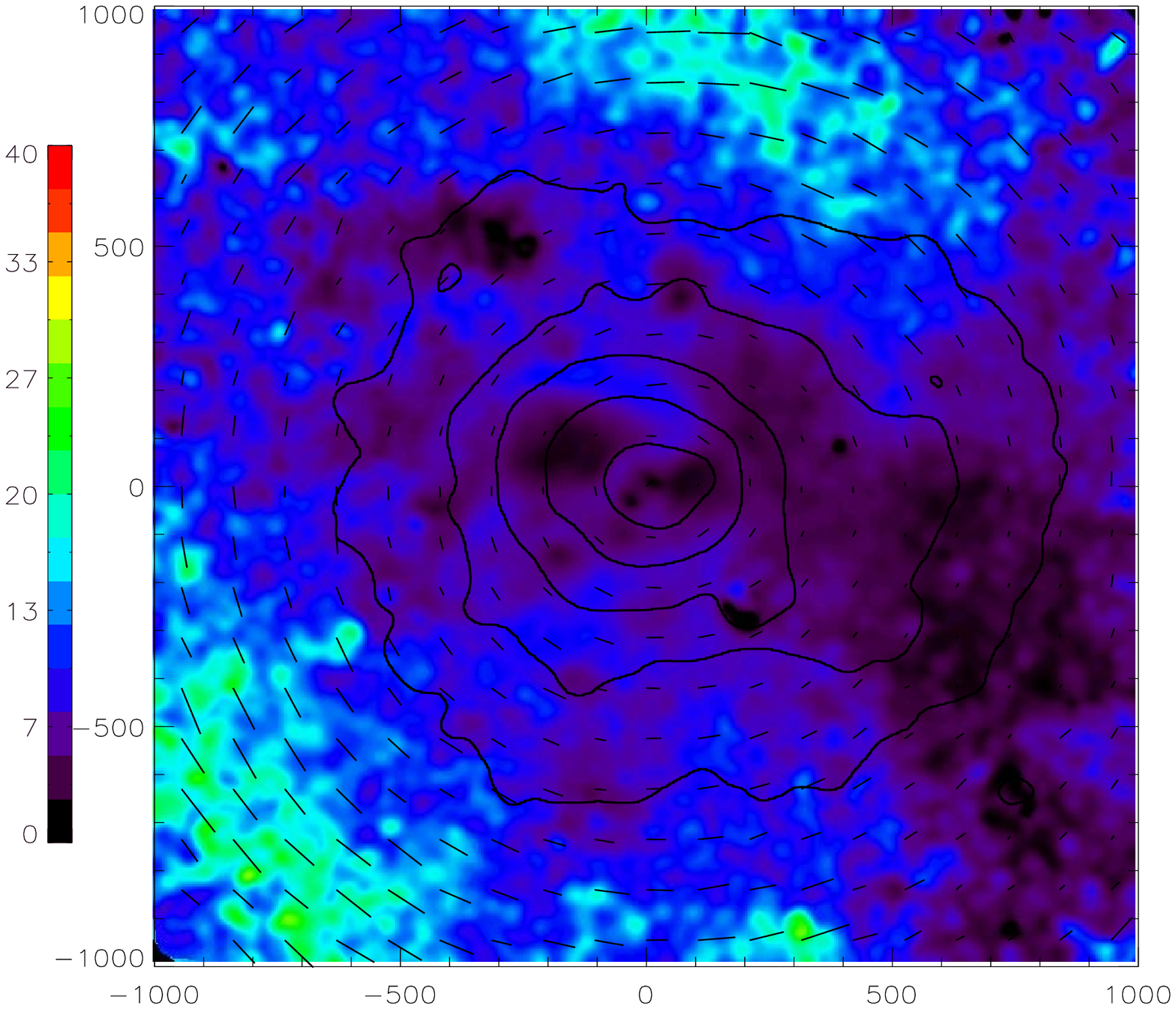}
\caption{Polarization degree of the simulated massive cluster g8
  \citep{Dol08} in the K$_\alpha$ line of Fe XXV at 6.7 keV. The
  polarization degree was evaluated as $P=\sqrt{Q^2+U^2}/I$. $I$ is
  the total intensity, including scattered and direct emission. The
  colors in the images denote the polarization degree in per cent. The
  short dashed lines show the orientation of the electric
  vector. Contours (factor of 4 steps in intensity) of the X-ray
  surface brightness in the chosen line are superposed. The size of each
  picture is 2$\times$2 Mpc. The left panel shows the case of gas
  being at rest. The right panel shows the case of the gas velocities
  obtained in the simulations.} 
\label{fig:3Dpol}       
\end{figure*}

\cite{Zhu10a}, used full 3D models of galaxy clusters taken from 
cosmological simulations \citep{Dol08, Spr01} to calculate the polarization 
degree with an account for gas motions. For one of
the most massive simulated clusters (g8 in the sample of
\citet{Dol08}) the polarization degree in the He-like iron line reaches
$\sim$ 25 per cent at a distance of $\sim$ 500 kpc from the center if
no turbulent motions are present (Fig. \ref{fig:3Dpol}, the left
panel). Inclusion of gas motions substantially decreases the
polarization down to $\sim$ 10 per cent (Fig. \ref{fig:3Dpol}, the
right panel) within the same region. They also made calculations for a 
bullet-like cluster ("g72" cluster
in the simulated sample of \cite{Dol08}), showing that the
polarization degree is less that 7 per cent within the region where
interaction of two sub-clusters occurs.

When polarization in lines is considered, one should take into account the 
contamination of the polarized flux in the resonant line by the 
unpolarized emission in the neighboring spectral continuum and nearby 
lines. The effect is severe. For example, for the 6.7 keV line an energy
resolution of $\sim$50 eV leads to a factor of 2 drop in the degree of
polarization. Also, at large distances from the cluster center
unpolarized CXB emission starts to dominate. 

Currently new concepts of X-ray polarimeters are under discussion 
\citep[see][e.g.]{Mul09, Sof08, Cos08}. Galaxy clusters are promising
but challenging targets for future polarimetric observations.  
Polarimeters should have good energy resolution (better than  100 eV), 
which is hardly reachable with the nearest future X-ray
polarimeters. In addition, arcminute angular resolution is needed to
resolve nearby clusters. Naturally, a large field of view and
effective area are required to collect large numbers of photons
\citep[see][]{Zhu10a}. 

\section{Impact on the line shape}
\label{sec:line}

\begin{figure*}
  \includegraphics[width=0.5\textwidth]{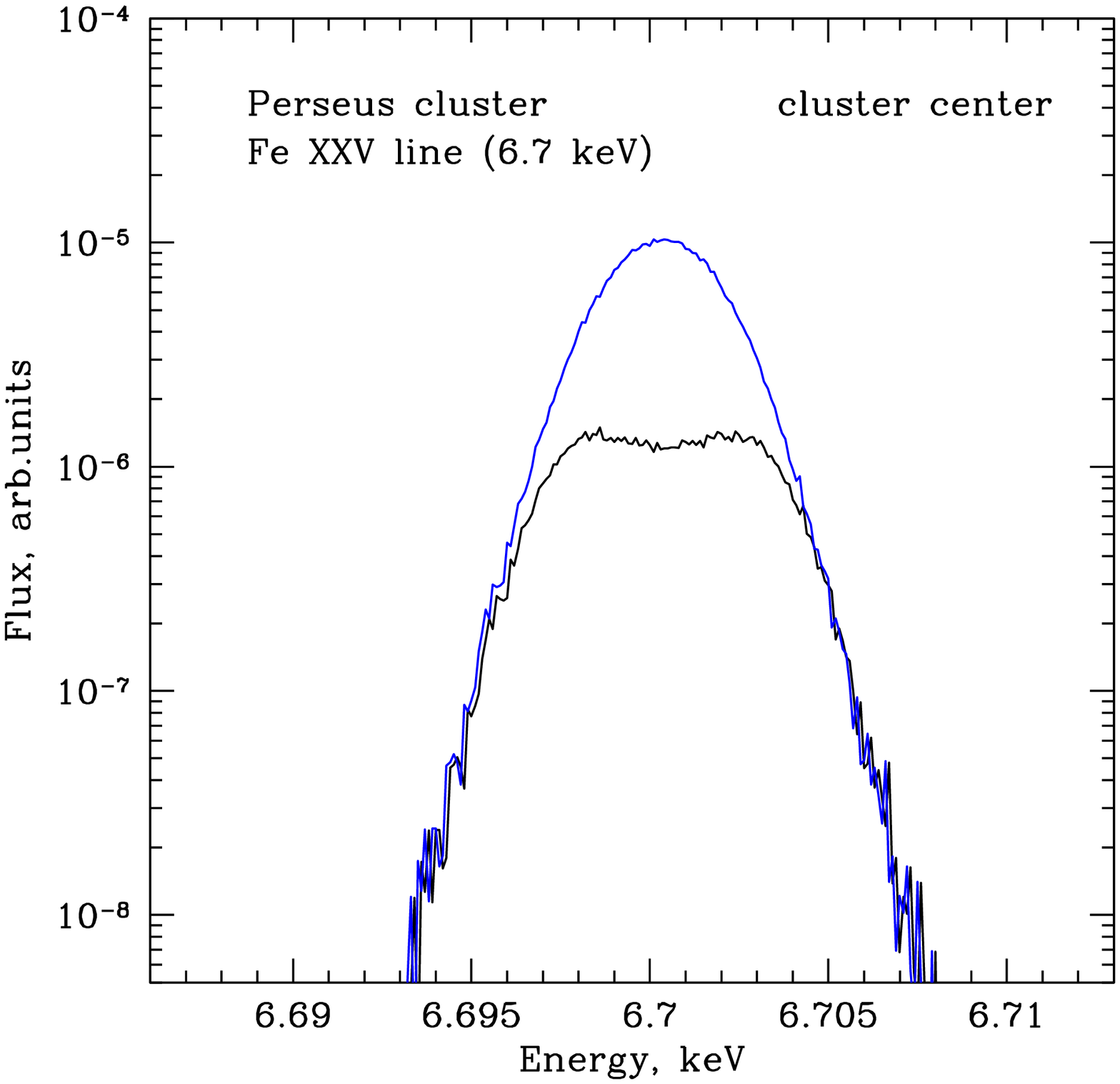}
  \includegraphics[width=0.5\textwidth]{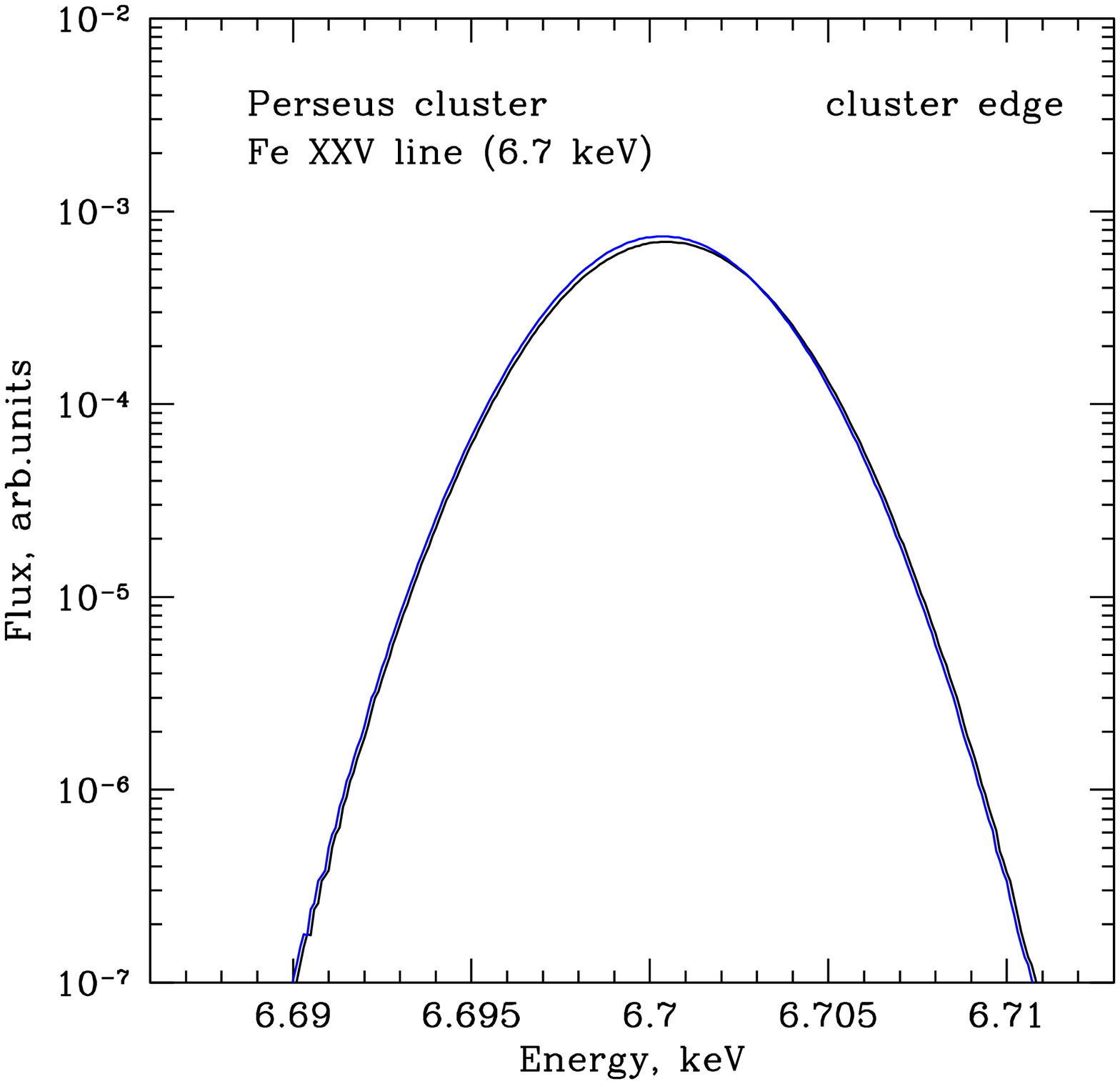}
\caption{Simulated spectra of the 6.7 keV line emerging from the core of
  the Perseus cluster (left) and at a projected distance of 800
  kpc (right). Lines are broadened only due to thermal motions of ions.
  The calculations were done with (black curves) and without (blue curves)
  accounting for resonant scattering. The line in the cluster
  outskirts (right panel) is broader due to the higher temperature in
  the outer regions of the Perseus cluster.} 
\label{fig:Perspec}       
\label{fig:line}       
\end{figure*}

The optical depth given by eq. \ref{eq:tau_0} corresponds to the center
of the line, i.e. to a photon energy equal to the line energy
$E_0$. If a photon is born in the wing of the line, then the optical 
depth will be lower,
\be
\tau(E)=\tau_0 e^{-\left (\frac{E-E_0}{\Delta E_D}\right )^2}.
\ee
If $\tau(E)\ll 1$, then such photons will leave the cluster without
scattering, while photons with $E\sim E_0$ will be scattered from the
line of sight going through the cluster center. It is therefore clear
that resonant scattering should affect the shape of optically  
thick lines. The spectral profiles of the He-like iron line at 6.7 keV
calculated in the core of the Perseus cluster and at a projected distance
of $\sim 800$ kpc are shown in Fig. \ref{fig:Perspec}. The calculations
were done with (black curves) and without (blue curves) allowance for
scattering. The lines are broadened only due to thermal motions of
ions. One can notice the suppression of the emission in the line
center (left panel) caused by resonant scattering: photons with the
energy close to the line central energy are scattered from the line of
sight because of the large optical depth, while photons born in the
wings escape freely. At the edge of the cluster the optical depth is
small and the line profiles almost match each other.

\section{Various extensions}
\label{sec:ext}

\subsection{Cosmological tests}
\label{sec:h0}
The effects of resonant scattering are proportional to the optical depth
of the cluster (see eq. \ref{eq:tau}), which scales linearly with the
size of the cluster $l$ and the gas density $n_e$: $\displaystyle
\tau\propto n_el\times F(T,Z)$, where the factor $F(T,Z)$ accounts for
the element abundance, ionization equilibrium and the line broadening. If
another characteristic of the cluster (with a different dependence on
$l$ and $n_e$) is known, then both $l$ and $n_e$ can be determined
simultaneously. For instance, one can use the X-ray surface brightness,
which is proportional to $\displaystyle n_e^2l\times W(T,Z)$, where
the $W(T,Z)$ factor accounts for the temperature dependent emissivity and
the gas metallicity. Once $l$ of a spherically symmetric cluster is
known, one can immediately perform an angular diameter-redshift test,
analogously as is done using a combination of the cosmic microwave
background distortions (the Sunyaev-Zel'dovich effect) and the X-ray surface
brightness. Thus a simple cosmological test based solely on X-ray data
is possible. This is an attractive possibility since such a test can
be performed by a single X-ray telescope in a single observation.

Several flavors of the same technique have been
proposed. \citet{1988ApJ...335L..39K} and \citet{1989ApJ...345...12S}
suggested to look for absorption lines in the spectrum of an X-ray
emitting background quasar.  The equivalent width of the strongest
absorption lines can be of order several eV. Another possibility is to
use the distortions of the cluster emission itself, using the polarization
\citep{Saz02} or shape of emission lines
\citep{2006ApJ...643L..73M}, or directly the radial variations of the 
equivalent width or the flux ratios of optically thick and thin lines
\citep[as discussed by][]{Gil87}. Some variants of these methods are
within the reach of currently operating X-ray telescopes, but the
uncertainties in line widths and the complex spectra of the central
regions of cool core clusters complicate the analysis. Significant
progress is expected with the launch of X-ray micro-calorimeters with
energy resolution of a few eV \citep[e.g.][]{Mit08}. They will
simultaneously constrain the presence of stochastic gas motions and
allow for accurate measurements of line shapes (see e.g. Fig.\ref{fig:line}).
 
\subsection{AGN X-ray echo in lines}
\label{sec:agn}

To better understand how massive black holes grow in galactic nuclei, 
one would like to have information on AGN variability on
time scales much longer than the several tens of years probed by
historic light curves. While the behavior on the longest time scales
(hundreds of millions of years) is determined by the supply
of gas (e.g. triggered by a merger) from the outer regions of a
galaxy to the central kiloparsec, AGN can switch on and off on intermediate
time scales for a number of reasons, e.g. due to instabilities in the
accretion disk or due to feedback of the growing massive black hole on the
surrounding medium. Clearly, such long-term phenomena can only be studied by
means of indirect observations. An attractive idea is to observe X-ray
radiation which was emitted in the past by a galactic nucleus and
later scattered in our direction by ambient gas
\citep{2002A&A...393..793S}. Cluster dominant elliptical galaxies are
the most promising targets for such observations, since their extended
gas haloes make it possible to probe time scales up to a few million
years. Since AGN radiation is much harder than the thermal emission of 
intracluster gas, one possibility is to search for AGN ``echoes''
at energies $E\gg kT/(1+z)$ ($T$ being the gas temperature and $z$ the
redshift of the object). An additional significant advantage is
provided by resonant X-ray lines, since the scattered/intrinsic flux
ratio is expected to be larger by a factor of 3--10 in a resonance
line than in the neighboring continuum. 

\begin{figure*}
  \includegraphics[width=0.5\textwidth]{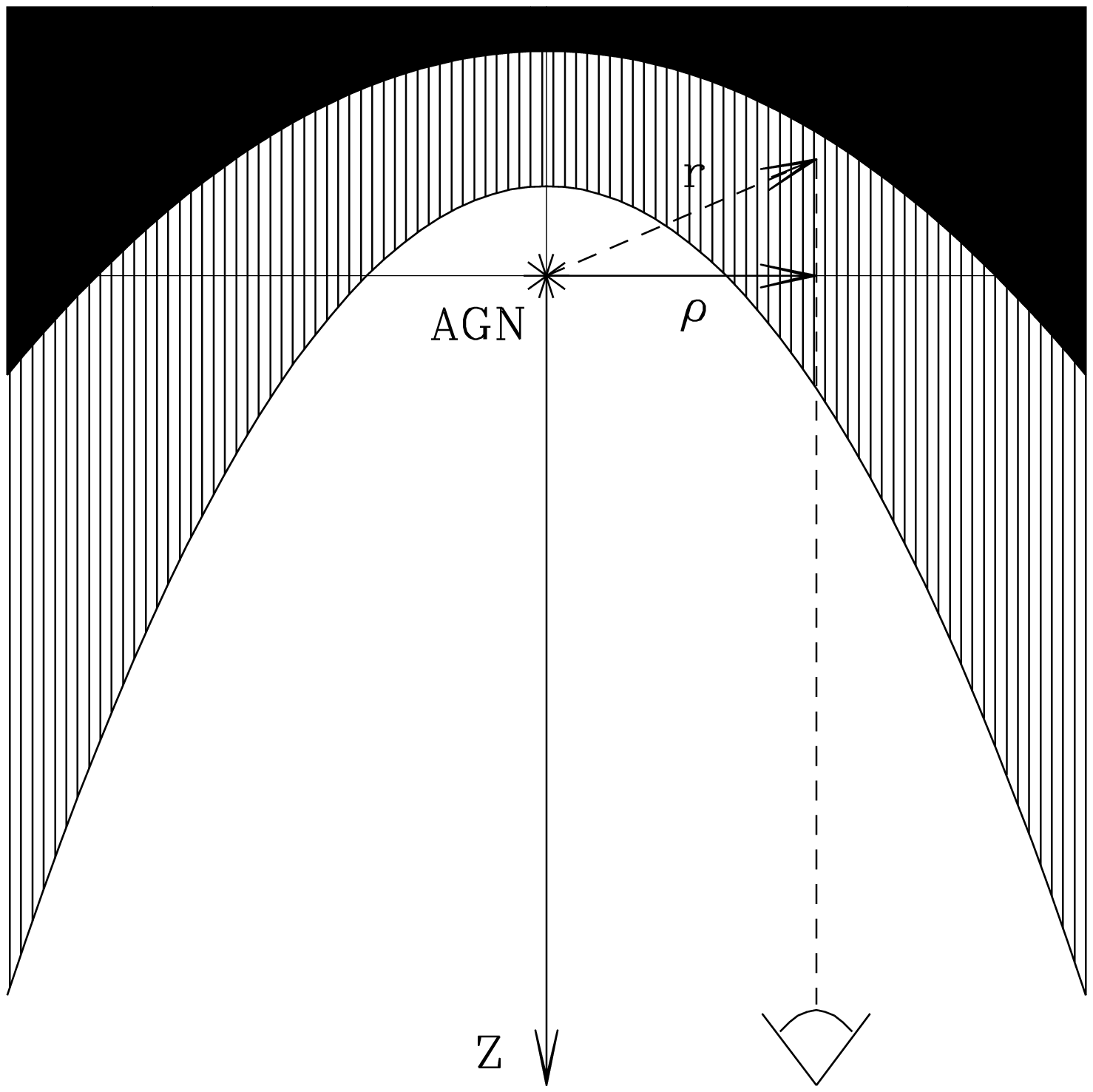}
  \includegraphics[width=0.5\textwidth]{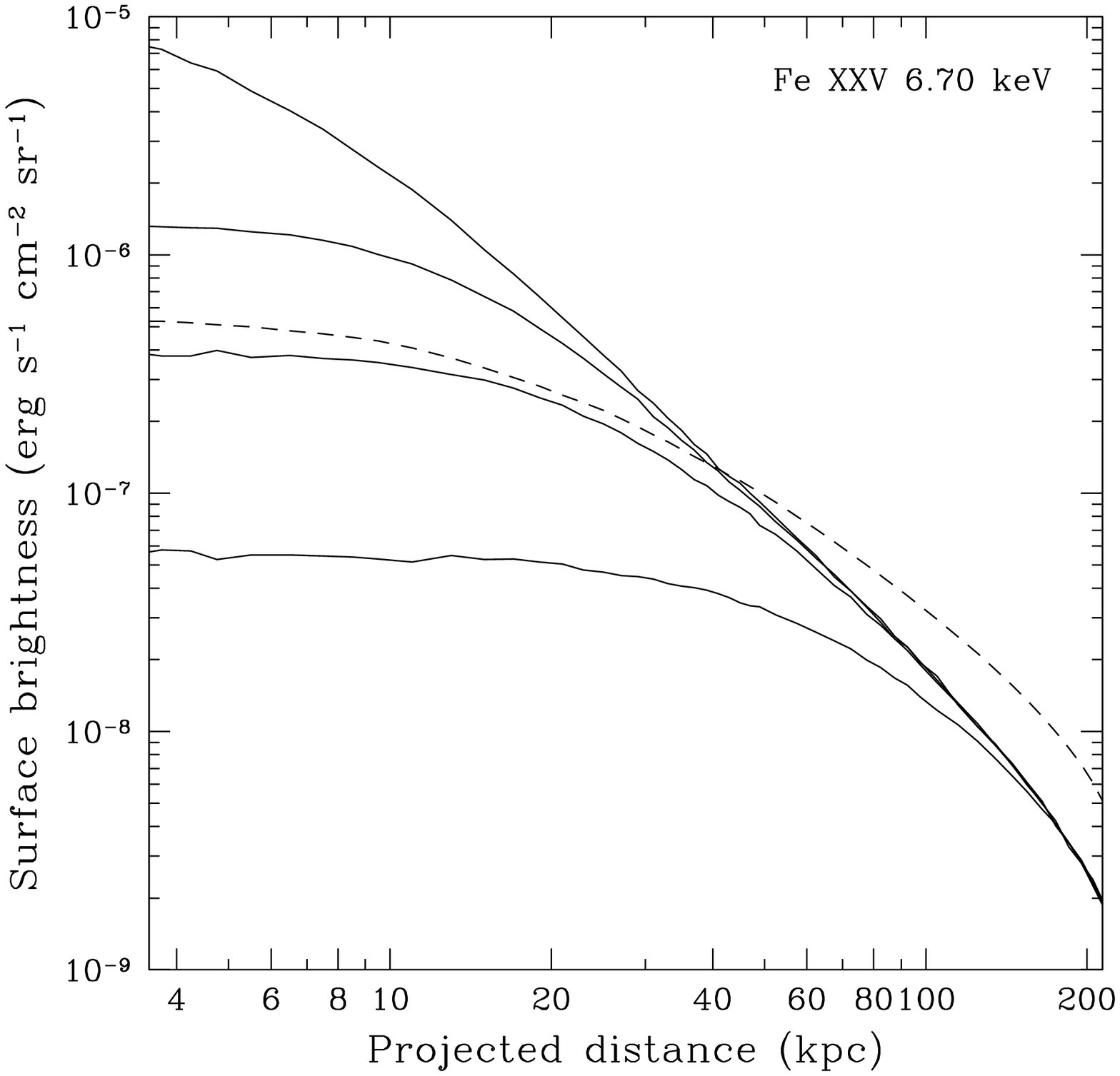}
\caption{{\bf Left:} Illustration of the delayed X-ray line echo of an AGN
outburst in a galaxy cluster. Locations with equal travel time from 
the AGN to the observer lie on a paraboloid. In the switch-off case
(the AGN was bright for a long time until some instant in the recent
past), the cluster plasma located in the grey and black regions is scattering 
the X-ray radiation emitted by the AGN. In the case of a short AGN flare,
only the grey area is filled by photons and contributing to the
scattered radiation. A possible path for a photon scattered into the
observer's line of sight is marked by a dashed line. {\bf Right:} 
M87/Virgo surface brightness of scattered AGN radiation (solid
lines) in the 6.70~keV line of He-like iron at times (from top to
bottom) of 5, 50, 100 and 250 thousand years after an AGN
switch-off. The dashed line shows the intrinsic surface brightness
profile of the cluster gas in the 6.70~keV line. The AGN luminosity
was $L_X=0.01~L_{Edd}$ before switch-off. Adapted from
\cite{2002A&A...393..793S}.}
\label{fig:agn}       
\end{figure*}

\cite{2002A&A...393..793S} assessed the level of constraints that
could be derived from future observations on the past X-ray luminosity
of M87 and Cyg~A (see Fig.~\ref{fig:agn}), giant elliptical galaxies
residing in the central regions of two nearby clusters of
galaxies. For instance, scattered line radiation should be detectable
from the Virgo cluster if the X-ray luminosity of M87 was a few
$10^{44}$~erg~s$^{-1}$ (i.e. three orders of magnitude brighter than
presently) until a few $10^5$~years ago. The same method can also be
applied to groups of galaxies 
and isolated giant elliptical galaxies, which too are large reservoirs
of hot gas capable of scattering AGN radiation. Of course, the
testable time scales are shorter (up to a few times $10^5$~years) than
for clusters of galaxies.

\subsection{Warm-Hot Intergalactic Medium (WHIM) and blazars}
\label{sec:whim}

\begin{figure*}
  \includegraphics[width=0.5\textwidth]{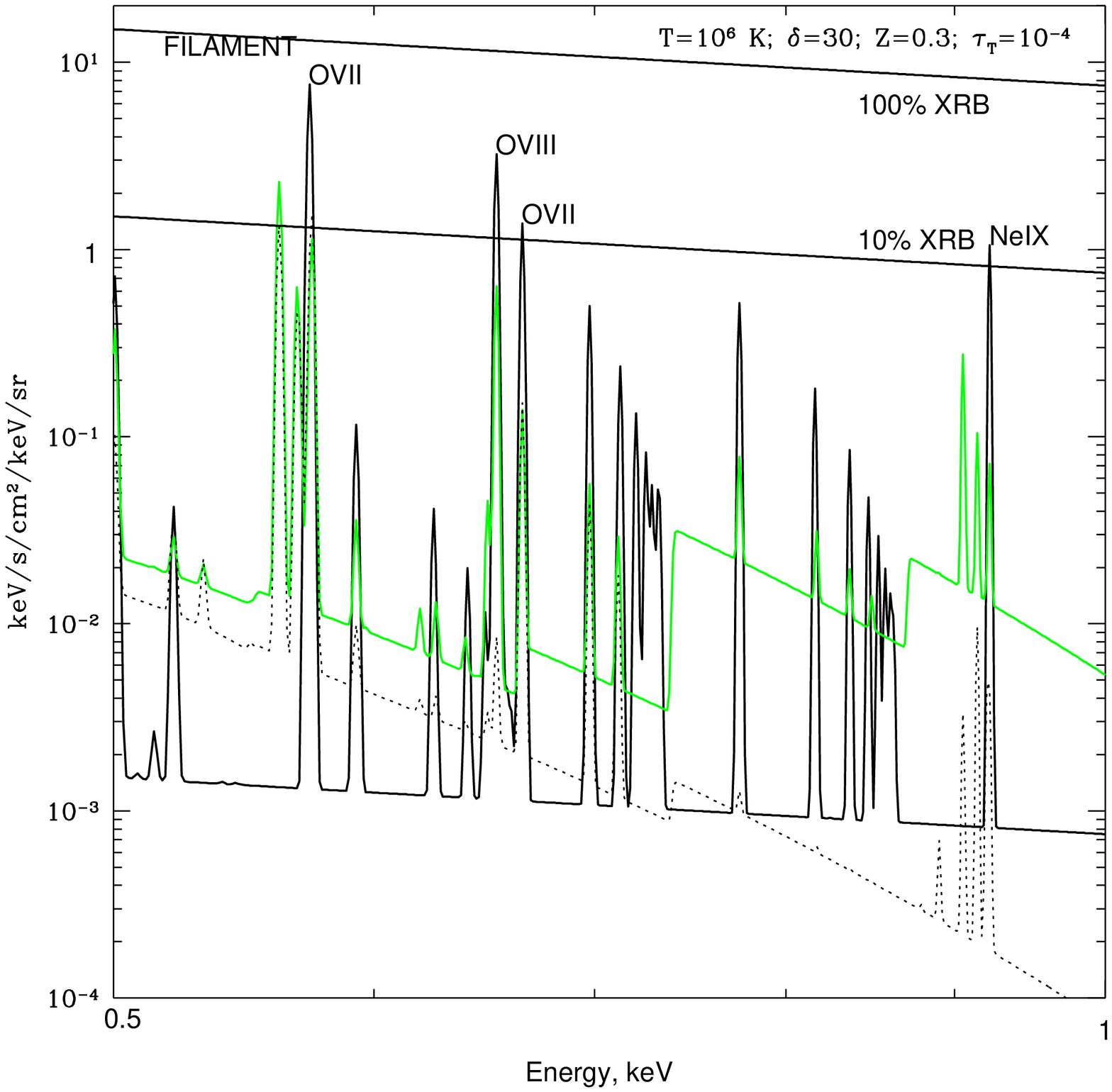}
  \includegraphics[width=0.5\textwidth]{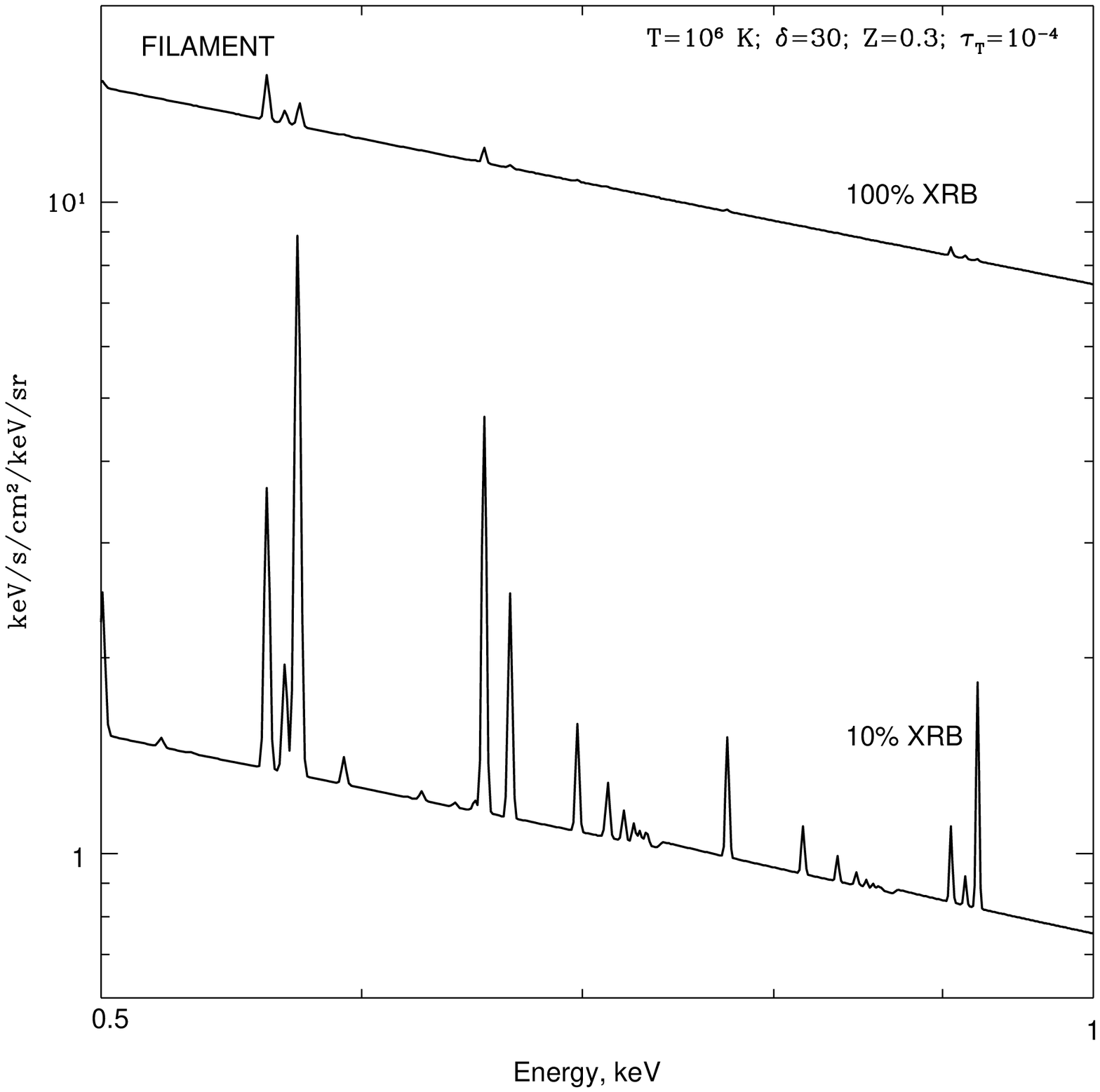}
\caption{{\bf Left: } The X-ray spectrum of warm gas with an
  overdensity of $\delta=30$, a temperature of $10^6$K, a Thomson
  optical depth of $10^{-4}$ and a metallicity of 30\% solar.  These
  conditions should be typical for a filamentary structure in the warm
  IGM. The dotted curve shows the emission spectrum due to
  collisional excitation and ionization. For the green curve
  photoionization by the CXB was taken into account. The black solid
  curve shows Thomson plus resonantly scattered radiation.  The upper
  solid line is the intensity of the CXB from the same region. The
  lower solid line shows 10\% of the CXB intensity (the level to which
  discrete sources may be removed). The spectra were convolved with a
  Gaussian with a FWHM of 2 eV. {\bf Right: } The upper curve shows
  the total spectrum of a region of the sky containing a filament with
  Thomson optical depth of $10^{-4}$. This spectrum is a sum of the
  thermal emission of the filament, scattered CXB and emission of the
  background sources, part of which is either absorbed or scattered by
  the filament. Note that resonance scattering does not change the CXB
  flux and only emission lines due to the thermal emission of the
  filament are visible in this spectrum.  The lower curve shows the
  spectrum of the same region if 90\% percent of the background is
  resolved into compact sources and removed. Note that the most
  prominent OVII line at 0.57 keV becomes visible only in this
  spectrum. In the hypothetical case that all CXB is resolved and
  removed the residual spectrum should be equal to the sum of the
  thermal emission of the filament and scattered CXB emission. These
  two components are shown in the left panel. Adapted from
  \citet{2001MNRAS.323...93C}.}
\label{fig:whim}       
\end{figure*}

Resonant scattering may be important not only in the dense gas inside
a cluster, but also in a much more tenuous warm intercluster medium in
filaments. In the local Universe such structures have linear sizes of
5--10 Mpc, temperatures of $\sim 10^5$ to $\sim 10^7$ K and density
$\delta=10-100$ times larger than the mean baryonic density 
of the Universe ($\rho_{\rm bar}= 3.6~10^{-31}~{\rm g~
  cm^{-3}}$). Simulations suggest that these structures may account
for a significant or even dominant fraction of the baryons at $z=0$
\citep[e.g.][]{1999ApJ...514....1C}. 

The low density of filaments 
makes direct detection of their thermal emission extremely
difficult. If there is a bright QSO then one can try to search for
absorption lines using high resolution spectroscopy
\citep[e.g][]{2005Natur.433..495N}, although it is not clear if
existing data have already provided a robust detection of truly diffuse
gas \citep[e.g][]{2007ApJ...656..129R}. If one is looking for WHIM in the 
directions where no bright QSO is present, then one can still use a 
combined signal of numerous background sources (and search for an absorption
feature) or instead look ``between'' the background sources. As
pointed out by \citet{2001MNRAS.323...93C} the 
scattering of X--ray background photons in He and H--like ions
of heavy elements can exceed the ``local'' thermal emission of a filament by a
factor of a few or more. Due to the conservative nature of 
resonant scattering, this resonantly scattered radiation can only be
identified if a significant fraction of the CXB is resolved and
removed. While the combined spectrum of the resolved sources will contain X--ray
absorption features, the residual background will contain
corresponding emission features with the (on average) same intensity.  At the
relevant densities and temperatures, the lines of He and H--like oxygen
at 0.57 and 0.65 keV are most promising. These lines (which have a
typical width of $\sim$ 1--2 eV) may contain up to 50\% of the total
0.5--1 keV emission of the filament (Fig. \ref{fig:whim}). Note that
the detection of WHIM in absorption using a combined signal of background
objects and the detection of a line in emission using a part of the
image without bright sources impose different requirements on the
telescope characteristics. In the former case, the energy resolution is
crucial, since the goal is to detect a weak absorption line in the
combined spectrum of all resolved sources -- in practice the best
results will be achieved if the energy resolution is better than the
equivalent width of the absorption line. In the latter case, the same
total flux is compared with the unresolved background -- the low
internal background is crucial in this case, while superb energy
resolution is less of an issue. In both cases the angular resolution has to
be good enough to resolve $\ge 50$\% of the CXB.

On average up to a few percent of the soft CXB could be resonantly
scattered by this phase of the IGM and resonantly scattered photons
should account for a significant fraction of the truly diffuse
background at low energies.  Close to bright X-ray sources such as galaxy
clusters or AGN the flux of scattered radiation will be further
enhanced.  From this point of view, off-line blazars are the most
promising illuminating sources. The scattered emission from AGN may
also constrain the duration of the active phase of these objects
analogously to what is discussed in \S\ref{sec:agn}.

\section{Conclusions}
Resonant scattering in the brightest X-ray emisson lines leads to a
number of characteristic observational features, inluding distortions
in the surface brightness, modifications of the line spectral shape
and polarization. The sensitivity of the resonant scattering effects to
the gas velocity field makes them a promising tool for IGM studies. Resonant
scattering can also be used for the basic cosmological angular
diameter/redshift test, as a tracer of a very rarefied IGM and as a
diagnostic of past outbursts from AGN. While the distortions in line 
surface brightness profiles are within the reach of modern X-ray
telescopes, future high energy resolution and polarimetric missions 
will be able to fully exploit resonant scattering as a diagnostic
tool.

\begin{acknowledgements}
We are grateful to Klaus Dolag, William Forman, Nail Inogamov, Dmitry
Nagirner, Leonid Vainshtein and Norbert Werner for useful discussions.
\end{acknowledgements}

\end{document}